\documentclass[a4paper,11pt]{article}
\usepackage{jcappub}
\usepackage{bm}
\usepackage{makecell}

\usepackage{graphicx}
\usepackage{newtxmath,newtxtext,amsfonts}
\usepackage[caption=false]{subfig}

\usepackage{textcomp}

\usepackage{float}

\usepackage{hyperref}
\hypersetup{
	colorlinks=true,
	linkcolor=blue,
	filecolor=violet,     
	urlcolor=blue,
	citecolor=red
}
\usepackage[]{natbib}
\usepackage{xcolor}
\usepackage{orcidlink}
\usepackage{epsfig}
\usepackage{caption}

\usepackage{commath}
\usepackage{cancel}
\usepackage{csquotes}
\usepackage{placeins}
\usepackage{graphicx}

\usepackage{float}

\usepackage{multirow}
\usepackage{tcolorbox}

\usepackage{orcidlink}

\begin{document}

\title{Constant-Roll Inflation as a Competitive Alternative to Slow-Roll Inflation}
\author[a]{Sandip Biswas\note{Corresponding author.}}
\author[b]{Rajib Saha,}
\author[a]{Kaushik Bhattacharya}

\affiliation[a]{Department of Physics, Indian Institute of Technology, Kanpur, Kanpur-208016, Uttar Pradesh, India}
\affiliation[b]{Department of Physics, Indian Institute of Science Education and Research, Bhopal, Bhopal By-Pass Road, Bhauri Bhopal, 462066
Madhya Pradesh, India}

\emailAdd{sandipb20@iitk.ac.in}
\emailAdd{rajib@iiserb.ac.in},
\emailAdd{kaushikb@iitk.ac.in}
\abstract{
Starting from the
constant-roll condition, which involves the constant-roll parameter $\beta$, which is a fundamental constant in the constant-roll inflationary models,
we derive the scalar power
spectrum and obtain an analytic expression for the scalar spectral index
$n_s$ in terms of $\beta$ and the number of remaining e-folds
$\tilde N$.  The resulting relation is non-monotonic in $\beta$,
allowing two distinct constant-roll branches, one with $\beta<0$ and one
with $\beta>0$, to reproduce nearly the same observed scalar tilt.
We constrain $\beta$ using Planck 2018 CMB data, and further using the
combined Planck 2018 + ACT DR6 data.  For Planck 2018, the posterior
distribution of $\beta$ exhibits a bimodal structure associated with the two
analytical branches, while the combined Planck+ACT analysis tightens the
allowed range of $\beta$ and shows continued consistency with the
$\beta=0$ slow-roll-like spectral point.  The data constrain $\beta$ to be
small, but do not require it to vanish exactly.  We also estimate the
inflationary mass scale $M$ by using the observed scalar-amplitude
normalization at the pivot scale.  Both the negative and positive $\beta$
branches give a high-scale inflationary normalization of order
$M\sim10^{-5}M_{\rm Pl}$. We show that the $\beta \to 0$ limit is only spectrally slow-roll like, but does not reproduce standard slow-roll dynamics. Our results show that constant-roll inflation is
observationally competitive with the usual slow-roll inflationary paradigm. The constant roll inflation model is supported by the latest CMB observations equally well or marginally better when
compared with the slow roll inflation model.}
\date{}

\maketitle

\section{Introduction}

Inflation is one of the most successful ideas in modern cosmology. It explains why the observable Universe is nearly homogeneous and isotropic on large scales, and it provides a natural mechanism for generating the primordial density perturbations that later seed the formation of cosmic structure. The simplest and most widely studied realization of inflation is slow-roll inflation \cite{Kazanas:1980tx,Guth:1980zm,Sato:1980yn,Sato:1981ds,Linde:1981mu,Riotto:2002yw}, where a scalar field evolves slowly along a sufficiently flat potential. This framework has been remarkably successful in explaining the observed nearly scale-invariant scalar power spectrum and the small red tilt measured in the cosmic microwave background~\cite{Guth:1982ec,Bardeen:1983qw,Mukhanov:1985rz,Sasaki:1986hm,Mukhanov:1981xt,Hawking:1982cz}.

Despite this success, slow roll is not the only possible way to realize inflation. The slow-roll approximation assumes that the acceleration of the scalar field is negligible compared with the Hubble friction term. This assumption is powerful and often well justified, but it is not a fundamental requirement for accelerated expansion. More general inflationary phases can occur when the scalar field acceleration is not negligible. Constant-roll inflation (CRI) \cite{Motohashi_2015,Motohashi_2017,Motohashi_20171,Motohashi_2017pbh,Yi:2017mxs,Gao_2017,Gao_2018,Gao_2019, Guerrero:2020lng, Cicciarella_2018,Karam_2018,Anguelova_2018,Odintsov_2017,Odintsov_2017_02,Mun_2021,Martin:2012pe,Gao_2019,Ito_2018,Carrasco_2015} provides a particularly simple and controlled way of studying such departures from slow roll. In this framework, the acceleration of the inflaton is not set to zero, but is instead assumed to remain proportional to the Hubble friction term, with the proportionality measured by a constant parameter usually denoted by $\beta$. In CRI models one therefore gets a new fundamental constant, $\beta$, which specifies the amount of constant-roll the system permits. 

The constant-roll parameter $\beta$ measures the departure from the standard slow-roll regime. When $\beta$ is close to zero, the usual slow-roll description is recovered. Nonzero values of $\beta$, on the other hand, describe inflationary dynamics in which the inflaton acceleration plays a more active role. This makes constant-roll inflation an interesting phenomenological extension of the standard slow-roll picture. It allows one to ask whether current cosmological observations truly require the inflaton to be in an almost slow-roll state, or whether small but finite constant-roll deviations are still allowed.

An important feature of constant-roll inflation is that it is not merely a phenomenological deformation. Once the constant-roll condition is imposed, the allowed form of the Hubble parameter as a function of the scalar field becomes highly constrained. 
From the allowed forms of the Hubble parameter one in general reconstructs the inflaton potential which can give rise to CRI.
Depending on the sign of $\beta$, the theory naturally separates into different branches. Positive and negative values of $\beta$ lead to different reconstructed scalar potentials, but these different potentials can nevertheless produce very similar scalar spectral indices over the observationally relevant range of scales. This opens the possibility of a branch degeneracy: two different constant-roll backgrounds may be observationally difficult to distinguish using current measurements of the scalar tilt alone.

A number of studies have been carried out over more than two decades to constrain the parameters of inflation such as scalar spectral index $n_{s}$, scalar-to-tensor ratio $r$, using CMB observations \cite{Leach_2002,Mortonson_2011,Easther_2012}. The purpose of this work is to confront this possibility with current CMB data. The cosmological parameters arising from inflationary theories serve as a fundamental probe to uncover the physics of the early universe. Satellite based CMB missions produced cosmological parameters with increasing precisions over the past several decades~\citep{CobeCosmo1992,WMAPCosmo2013, Aghanim2020}. These measurements have been augmented by small scale and ground based CMB observation by ACT collaboration~\citep{ACTCosmo2025}. Cosmological parameters were estimated from the simulated observations of future generation CMB experiments in the case of interacting dark matter and dark energy models by~\cite{MNRAS_Saha2}. In~\cite{APJ_Saha} authors estimate tensor to scalar ratio using simulated observations of future generation low noise CMB B mode observations. Cosmological parameters with a dark energy component evolving in an oscillatory tracker potential have been constrained by~\cite{Saha2022}. In~\cite{MNRAS_Saha1} perform a foreground model independent Bayesian CMB reconstruction method on simulated observations of CMB E mode and constrain  the relevant cosmological parameters. In~\cite{SahaSrikanta2022} the authors perform a direct estimation of the cosmological density parameters using Hubble parameter data.  A machine learning approach was implemented to measure cosmological density parameters and was compared with the results of the traditional MCMC approach by~\cite{SahaSrikanta2024}. 

{A number of studies has also been carried out by many author on constraining the parameters of constant roll inflation. The very first of which was \cite{Motohashi_20171}, where Motohashi and Starobinsky confronted the constant-roll inflationary scenario with observational data by using the exact background solution associated with the constant-roll condition $\ddot{\phi}=\beta H\dot{\phi}$. However, they evaluated the scalar spectral index and tensor-to-scalar ratio using the standard slow-roll relations $n_s-1=-6\epsilon_1+2\eta$ and $r=16\epsilon$. Using the Planck 2015 and {BICEP2/Keck} constraints, they obtained an observationally viable region corresponding approximately to $0.01\lesssim\beta\lesssim0.02$, indicating that the observationally allowed models lie close to the slow-roll regime. In \cite{Gao_2018} the author investigated observational constraints on constant-roll inflation by considering models with constant second hubble-flow parameter $\epsilon_2$ and a constant parameter $\bar{\eta}=\eta_H+\epsilon_1$. Using a Bessel-function approximation, analytical expressions for the scalar and tensor power spectra, the scalar spectral index $n_s$, and the tensor-to-scalar ratio $r$ were derived to first order in $\epsilon_1$. The resulting predictions were compared with the {Planck 2015} constraints in the $n_s-r$ plane. Yi and Gong~\cite{Yi:2017mxs} investigated the primordial perturbations in constant-roll inflation and examined the validity of the commonly used Bessel-function approximation for calculating the scalar and tensor power spectra. They derived analytical expressions for the scalar and tensor power spectra, as well as the corresponding spectral tilts and tensor-to-scalar ratio, up to first order in the Hubble slow-roll parameter $\epsilon_1$. They also studied the evolution of the curvature perturbation on super-horizon scales and found that it remains constant for the inflationary solution considered. By comparing the analytical results with numerical solutions of the Mukhanov--Sasaki equations, they showed that the Bessel-function approximation agrees well with the numerical results when the constant-roll parameter $\eta_H$ is small, but can deviate significantly on super-horizon scales when $\eta_H$ is not small. They further found that the particular constant-roll model considered is inconsistent with the {Planck 2015} constraints at the $68\%$ confidence level. }

In the present paper, using CMB data we derive the scalar perturbation spectrum for CRI and obtain the scalar spectral index in terms of the constant-roll parameter and the number of $e$-folds before the end of inflation.
The resulting relation shows that, for a fixed number of $e$-folds, the observed scalar tilt can correspond to two distinct values of the constant-roll parameter. In particular, for a representative value of sixty $e$-folds, one finds one branch with negative $\beta$ and another branch with positive $\beta$. These two branches correspond to different constant-roll potentials, but both can reproduce the observed scalar spectral index.

We then use Planck 2018 CMB data \cite{Aghanim2020} to constrain the constant-roll parameter. The Planck-only analysis shows that the posterior distribution of $\beta$ has a two-branch structure, reflecting the analytical degeneracy of the model. The data constrain $\beta$ to be close to the slow-roll limit, but they do not force $\beta$ to vanish exactly. Small positive and negative constant-roll values remain compatible with the observed scalar spectrum. We further include ACT DR6 data \cite{ACTCosmo2025}
 together with Planck, which tightens the allowed range of $\beta$ and begins to sharpen the structure of the posterior distribution.

Our results show that CRI is observationally competitive with the standard slow-roll description. When $\beta$ is small, the scalar spectral predictions approach a slow-roll-like form. However, the strict $\beta\to0$ limit of the
reconstructed constant-roll branches is not a generic rolling slow-roll potential limit, but rather a degenerate de Sitter limit. Thus the connection with slow roll is primarily spectral or phenomenological. Moreover, the best-fit comparison suggests that constant-roll inflation may provide a marginally improved fit to the current CMB data, although the improvement is not statistically decisive. Thus, the main conclusion is not that constant-roll inflation is preferred over slow roll, but rather that it remains a viable and competitive alternative that deserves further serious investigation.

We also estimate the characteristic energy scale associated with the allowed constant-roll branches. Using the observed amplitude of the scalar power spectrum, we find that both the positive and negative $\beta$ branches lead to an inflationary scale of the usual high-scale order, close to $10^{-5}M_{\rm Pl}$. This is a perfectly consistent scale of inflation accoring to modern cosmological observation \cite{Aghanim2020}. This agreement between the two branches further supports the interpretation that the branch structure is a genuine feature of the constant-roll dynamics rather than an artifact of an inconsistent normalization. Our analysis shows that the constant-roll parameter $\beta$ can be treated as a fundamental inflationary parameter. It controls the departure from the slow-roll attractor, classifies distinct inflationary branches, determines the scalar spectral tilt, and can be directly constrained by CMB observations. 

The paper is organized as follows. In the next section, we discuss the background dynamics of constant-roll inflation and reconstruct the scalar potentials associated with the positive and negative $\beta$ branches. In Section \ref{sec:perturbation}, we derive the scalar perturbation spectrum and the corresponding scalar spectral index. In Section \ref{sec:data_analysis}, we test our theoretical model predictions in light of the Planck 2018 and the Planck plus ACT DR6 data and obtain constraints on the constant-roll parameter. The analysis does not yield any information on the parameters present in the inflaton potential. In Section \ref{param} we find out the possible values of the parameters in the CRI potential by sampling the constant-roll parameter and the power spectrum at the pivot scale. In Section \ref{sll}, we point out the subtelety related to the slow-roll limit of CRI. In the last section, Section  \ref{conclu} we conclude the paper after presenting a brief summary of all the relevant points discussed in the paper.

\section{Background dynamics of CRI}\label{sec:Background dynamics of CRI}

The canonical single-field inflation is described by the action:
\begin{equation}
S=\int d^4x\sqrt{-g}\left[\frac{M_{\rm Pl}^2}{2}R-\frac12(\partial\phi)^2-V(\phi)\right]\,,
\end{equation}
where $\phi$ is the inflaton field, $V(\phi)$ is the inflaton potential and $R$ is the Ricci potential. For the spatially flat Friedmann-Lemaitre-Robertson-Walker (FLRW) background:
\begin{eqnarray}
ds^2=dt^2 - a^2(t)d{\bf x}^2\,,
\end{eqnarray}
where $a(t)$ is the scale factor and $H\equiv \dot{a}/a$ is the Hubble parameter, the equations of motion are
\begin{eqnarray}
3M_{\rm Pl}^2H^2 &=&\frac12\dot\phi^2+V(\phi)\,,\\
-2M_{\rm Pl}^2\dot H&=&\dot\phi^2\,,
\end{eqnarray}
and
\begin{equation}
\ddot\phi+3H\dot\phi+V_{,\phi}=0\,.
\end{equation}
In the above equations the dot specifies a derivative with respect to the cosmological time.

In slow-roll inflation, one imposes the slow-roll conditions on the above equations and then proceeds to solve for the background solution. Instead of imposing the slow-roll condition 
$|\ddot\phi|\ll |H\dot\phi|$, this analysis adopts the constant-roll condition:
\begin{equation}
\ddot\phi=-3\beta H\dot\phi\label{cr_cond},
\end{equation}
a framework introduced in\cite{Motohashi_2015}, where $\beta$ is a constant parameter. 
The slow-roll limit corresponds to $\beta\simeq 0$, whereas ultra slow-roll inflation \cite{} limit is for $\beta=1$. Using the Hamilton--Jacobi formalism with $H=H(\phi)$, one obtains:
\begin{equation}
\dot\phi=-2M_{\rm Pl}^2H_{,\phi}\label{eq_phidot}.
\end{equation}
Differentiating with respect to time and substituting the constant-roll condition yields
\begin{equation}
H_{,\phi\phi}=\frac{3\beta}{2M_{\rm Pl}^2}H \label{eq:H_diff_eq}\,.
\end{equation}
The subscript, comma followed by $\phi$, stands for a derivative with respect to the inflaton field $\phi$. The structure of the solution, of the above equation, depends on the sign of $3\beta$. Through the standard Hamilton--Jacobi relation, the inflationary potential can be reconstructed from \cite{}:
\begin{equation}
V(\phi)=3M_{\rm Pl}^2H^2-2M_{\rm Pl}^4H_{,\phi}^2\label{eq_V}.
\end{equation}
This is the general form of the inflaton potential, which can appear in CRI, given by the Hamilton-Jacobi equation and the constant-roll condition which produces a form of $H(\phi)$ which yields the above form of the potential. 

For $3\beta>0$, the general solution is spanned by a linear combination of exponential modes:
\begin{equation}
H(\phi)=C_1\exp\left(
\sqrt{\frac{3\beta}{2}}
\frac{\phi}{M_{\rm Pl}}
\right)
+
C_2\exp\left(
-\sqrt{\frac{3\beta}{2}}
\frac{\phi}{M_{\rm Pl}}
\right)\label{eq:H_pb_sol},
\end{equation}
where $C_{1}$ and $C_2$ are arbitrary integration constant. Restricting to the sector $C_1 C_2 > 0$, we map these degrees of freedom onto a single hyperbolic trajectory by defining the mass scale $M \equiv 2\sqrt{C_1 C_2}$ and the field shift $\phi_0 \equiv \frac{1}{2}\sqrt{\frac{2}{3\beta}} M_{\rm Pl} \ln(C_1/C_2)$. This transforms the solution into:
\begin{equation}
H(\phi)=M\cosh\left(
\sqrt{\frac{3\beta}{2}}
\frac{\phi+\phi_0}{M_{\rm Pl}}
\right)\label{eq_H_cosh},
\end{equation}
Here, $\phi_0$ acts as a translation parameter along the field axis, while $M$ sets the baseline Hubble scale. Setting $\phi_0 = 0$ recovers the symmetric branch used throughout the standard framework.

Substituting Eq.~\eqref{eq_H_cosh} into Eq.~\eqref{eq_V} we get the potential
\begin{equation}
V(\phi)=3M^2M_{\rm Pl}^2
\left[
1-\frac{(1-\beta)}{2}
\left\{
1-\cosh\left(
\sqrt{6\beta}
\frac{\phi}{M_{\rm Pl}}
\right)
\right\}
\right]\label{potential_positive_beta}.
\end{equation}
By substituting Eq.~\eqref{eq_H_cosh} into Eq.~\eqref{eq_phidot} and integrating, the corresponding exact background solution is found to be
\begin{equation}
\phi(t)=
M_{\rm Pl}\sqrt{\frac{2}{3\beta}}
\ln\left[
\coth\left(\frac{3\beta}{2}Mt\right)
\right]\label{eq_phi_t}\,.
\end{equation}
Substituting Eq.~\eqref{eq_phi_t} into Eq.~\eqref{eq_H_cosh} yields the explicit time dependence for the expansion rate:
\begin{equation}
H(t)=M\coth(3\beta Mt)\label{H_beta1}\,,
\end{equation}
which integrates directly to give the cosmic scale factor:
\begin{equation}
a(t)\propto \sinh^{\frac{1}{3\beta}}(3\beta Mt)\label{eq_a_t}\,.
\end{equation}
Alternatively, choosing $C_1 = -C_2 = M/2$ in Eq.~\eqref{eq:H_pb_sol} gives
\begin{equation}
H(\phi) = M \sinh\left( \sqrt{\frac{3\beta}{2}} \frac{\phi}{M_{\rm Pl}} \right). \label{eq:H_sinh_sol}
\end{equation}
Substituting Eq.~\eqref{eq:H_sinh_sol} into the constant-roll condition given in Eq.~\eqref{eq_phidot} and integrating the resulting differential equation yields the time evolution of the inflaton field:
\begin{equation}
\phi(t)=\sqrt{\frac{2}{3\beta}}\,M_{\rm Pl}\,\sinh^{-1}\!\left[\tan\!\left(C-3M\beta t\right)\right], \label{eq:phi_t_sinh}
\end{equation}
where $C$ is the integration constant. Substituting Eq.~\eqref{eq:phi_t_sinh} into Eq.~\eqref{eq:H_sinh_sol} yields the cosmic time dependence for the Hubble parameter:
\begin{equation}
H(t) = M\tan\left[ -3\beta M t + C \right]. \label{eq:H_t_sinh}
\end{equation}
As shown in Ref.~\cite{Biswas:2025vlz}, omitting temporal integration constants can lead to erroneous conclusions regarding model viability. Maintaining the constant $C$ in Eq.~\eqref{eq:H_t_sinh} is mathematically vital; keeping this shifting constant allows the argument to be localized within a physically stable quadrant to ensure $H(t) > 0$, cleanly accessing the inflationary window ($\ddot{a}>0$).
Integrating Eq.~\eqref{eq:H_t_sinh} gives the corresponding scale factor:
\begin{equation}
a(t) = a_0 \cos ^{\frac{1}{3\beta}}\left( -3\beta M t + C \right), \label{eq:a_t_sinh}
\end{equation}
where $a_0$ is the scale factor integration constant. The corresponding potential for CRI is given by 
\begin{equation}
	V(\phi) = 3M^2M_{\rm Pl}^2\left[\left(1-\beta\right)\cosh^{2}\left(\sqrt{{\frac{3}{2}\beta}}\frac{\phi}{M_{\rm Pl}}\right) - 1\right]\label{cr_v_beta_positive}.
\end{equation}
All the results presented till now require $\beta>0$. If $\beta<0$, one has to recast the calculation in a slightly different line.

For the case $\beta < 0$, we define $\tilde{\beta} \equiv -\beta > 0$. In this regime, Eq.~\eqref{eq:H_diff_eq} admits oscillatory solutions for the Hubble parameter as a function of the scalar field:
\begin{equation}
H(\phi)=
C_1\cos\left(
\sqrt{\frac{3\Tilde{\beta}}{2}}
\frac{\phi}{M_{\rm Pl}}
\right)
+
C_2\sin\left(
\sqrt{\frac{3\Tilde{\beta}}{2}}
\frac{\phi}{M_{\rm Pl}}
\right)\label{cr_h_beta_negative}\,,
\end{equation}
where $C_1$ and $C_2$ are integration constants. Similar to the case for $\beta>0$, we can rewrite Eq.~\eqref{cr_h_beta_negative} as
\begin{equation}
H(\phi)=
M\cos\left(
\sqrt{\frac{3\Tilde{\beta}}{2}}
\frac{\phi}{M_{\rm Pl}}
\right)\label{H_ne_beta}.
\end{equation}
The corresponding potential can be written as 
\begin{equation}
V(\phi)=
3M^2M_{\rm Pl}^2
\left[
1-\frac{(1+\Tilde{\beta})}{2}
\left\{
1-
\cos\left(
\sqrt{6\Tilde{\beta}}
\frac{\phi}{M_{\rm Pl}}
\right)
\right\}
\right]\label{V_for_betatilde}.
\end{equation}
The exact field solution corresponding to Eq.~\eqref{H_ne_beta} for the homogeneous background is
\begin{equation}
\phi(t)=
2M_{\rm Pl}
\sqrt{\frac{2}{3\Tilde{\beta}}}
\tan^{-1}(e^{3\Tilde{\beta}Mt})\label{phi_nb_t},
\end{equation}
with
\begin{equation}
H(t)=-M\tanh(3\Tilde{\beta} Mt)\label{H_nb_t},
\end{equation}
and
\begin{equation}
a(t)\propto
\cosh^{-1/3\Tilde{\beta}}((3\Tilde{\beta}) Mt)\label{a_nb_t}.
\end{equation}
Although Eq.~\eqref{H_nb_t} appears to give $H<0$ for $t>0$, the time coordinate is shifted such that the inflationary evolution occurs for $t<0$ \cite{Motohashi_2015}. In this convention, $t\to-\infty$ corresponds to $\phi\to0^+$ in Eq.~\eqref{phi_nb_t}, where $V$ in Eq.~\eqref{V_for_betatilde} is maximal, while $t\to0^-$ corresponds to the inflaton rolling down the potential toward $V\to0$.

We now introduce Hubble-flow parameters defined as 
\begin{equation}
    \epsilon_{n+1}\equiv \frac{\dot{\epsilon}_n}{H {\epsilon}_n}\label{eq_flow_params},
\end{equation}
with $\epsilon_0\equiv 1/H$. The next three members of the hierarchy are:
\begin{align}
    \epsilon_1\equiv-\dot{H}/H^{2},&&\epsilon_2\equiv\frac{\dot{\epsilon_{1}}}{H\epsilon_{1}} &&\text{and}&& \epsilon_3
\equiv
\frac{\dot{\epsilon}_2}{H\epsilon_2}\label{ep2_ep3_def}.
\end{align}
Using the definition of $\epsilon_1$, \(\epsilon_2\) can be expanded as 
\begin{equation}
    \epsilon_2=2\epsilon_1+2\frac{\ddot\phi}{H\dot\phi}\,.
\end{equation}
Employing  the constant-roll condition in Eq.~\eqref{cr_cond} yields
\begin{equation}
\epsilon_2=2\epsilon_1-6\beta
\label{ep2_ep1_rel}.
\end{equation}
Differentiating the above relation with respect to time, we obtain
\begin{equation}
\dot{\epsilon}_2=2\dot{\epsilon}_1.
\end{equation}
Assuming ${\epsilon}_2$ and ${\epsilon}_1$ are nonzero, substituting this result into the definition of $\epsilon_3$ from Eq.~\eqref{ep2_ep3_def}, we obtain
\begin{equation}
\epsilon_3
=
2\epsilon_1\label{ep3_ep1_rel}.
\end{equation}
These are the basic definitions of the various Hubble-flow parameters and their interrelations. 

Next, we recall the standard relation between the scale factor and the Hubble parameter,
\begin{equation}
\frac{a'}{a}=aH\label{eq:scale_factor_relation},
\end{equation}
where a prime denotes differentiation with respect to the conformal time $\tau$, defined as $a(\tau) d\tau\equiv  dt$. 
This allows us to rewrite the relation Eq.~\eqref{eq_flow_params} between higher order flow parameters
\begin{align}
   \epsilon_{n+1}&\equiv \frac{\dot{\epsilon}_n}{H {\epsilon}_n} 
   =\frac{{\epsilon}'_n}{H {\epsilon}_n}\frac{d\tau}{dt} 
   =\frac{{\epsilon}'_n}{aH {\epsilon}_n}\label{flow_parm_rel}\,.
\end{align}
To evaluate the first Hubble-flow parameter at the horizon exit of the pivot scale, we express $\epsilon_{1*}$ in terms of the remaining number of e-folds. The latter is defined as
\begin{equation}
\tilde N
\equiv
-\int_{t_e}^{t_*}H\,dt,
\end{equation}
which implies
\begin{equation}
d\tilde N=-H\,dt\,.
\end{equation}
The above minus sign occurs as we keep $t_e$ fixed but differentiate $\tilde{N}$ with respect to $t_*$. Here $dt$ is an infinitesimal change of $t_*$ and $\tilde{N} \ge 0$. Using Eqs.~\eqref{eq_flow_params} and \eqref{ep2_ep1_rel}, we obtain
\begin{equation}
\frac{d\epsilon_1}{d\tilde N}
=
-\epsilon_1\epsilon_2
=
-2\epsilon_1(\epsilon_1-3\beta)\,.
\end{equation}
Integrating from the horizon exit of the pivot scale, where $\epsilon_1\equiv\epsilon_{1*}$, to the end of inflation, indicated by $\epsilon_1\equiv\epsilon_{1e}=1$, yields
\begin{equation}
\int_{\tilde{N}}^0\tilde N
=
-\int_{\epsilon_{1*}}^{\epsilon_{1e}}
\frac{d\epsilon_1}
{2\epsilon_1(\epsilon_1-3\beta)}=-\frac{1}{6\beta}
\int_{\epsilon_{1*}}^{1}
\left(
-\frac{1}{\epsilon_1}
+
\frac{1}{\epsilon_1-3\beta}
\right)d\epsilon_1\nonumber\,.
\end{equation}
The above integration yields
\begin{align}
\tilde{N}
=
\frac{1}{6\beta}
\ln\left|
\frac{(1-3\beta)\epsilon_{1*}}
{\epsilon_{1*}-3\beta}
\right|\nonumber\nonumber\,.
\end{align}
Exponentiating both sides gives
\begin{equation}
e^{6\beta \tilde N}
=
\frac{(1-3\beta)\epsilon_{1*}}
{\epsilon_{1*}-3\beta}\nonumber.
\end{equation}
Finally, solving for $\epsilon_{1*}$ yields
\begin{equation}
\epsilon_{1*}
=
\frac{3\beta}
{1-(1-3\beta)e^{-6\beta \tilde N}}
\label{ep1_@_p.s}.
\end{equation}
This is the relevant expression for the Hubble-flow parameter $\epsilon_1$ when the pivot scale perturbation mode goes out of the horizon. Next, we discuss the properties of the perturbations generated in CRI and the scalar power spectrum related to it.

\section{Power spectrum in CRI}\label{sec:perturbation}

We now consider the evolution of scalar perturbations for the background solutions given in Eqs.~(\ref{potential_positive_beta}) and Eqs.~(\ref{V_for_betatilde}). 
The gauge-invariant comoving curvature perturbation $\zeta_k$ is related to the scalar metric perturbation through
\begin{equation}
 g_{ij}=-a^2(1-2\zeta)\delta_{ij},
\end{equation}
in the comoving gauge $\delta\phi=0$   
 \cite{Motohashi_2015,Mukhanov:1985rz,Mukhanov:1981xt,brandenberger1993classicalquantumtheoryperturbations}, where the metric in this gauge looks like:
\begin{equation}
    g_{\mu\nu}=a^2(\tau)\begin{pmatrix}
    (1+2\Phi) && -\partial_{j}B\\ 
    -\partial_{i}B && -(1-2\zeta)\delta_{ij}
    \end{pmatrix}\,,
\end{equation}
where $\Phi$ and $B$ are the other relevant functions in the present case. In the perturbation analysis we work in conformal time $\tau$. All background quantities obtained in the previous section as functions of cosmic time are now understood as functions of $\tau$ through the composition \(t=t(\tau)\). In particular, \(H(\tau)\equiv H(t(\tau))\) denotes the usual cosmic-time Hubble parameter. A prime denotes differentiation with respect to \(\tau\), whereas a dot denotes differentiation with respect to cosmic time \(t\). Introducing the Mukhanov--Sasaki (MS) variable $v_k\equiv M_{\rm Pl} z \zeta_k$,
with $z\equiv a\sqrt{2\epsilon_1}$,
the scalar perturbations satisfy the Mukhanov--Sasaki equation~\cite{Mukhanov:1985rz,Sasaki:1986hm}:
\begin{equation}
v_k''+\left(k^2-\frac{z''}{z}\right)v_k=0.
\label{MS_eq}
\end{equation}
To evaluate the evolution of the MS-variable, we differentiate $z$ with respect to conformal time:
\begin{equation}
z'=\sqrt{2}\left(a'\sqrt{\epsilon_1}+\frac{a}{2}\frac{\epsilon'_1}{\sqrt{\epsilon_1}}\right)\label{eq_z'}.
\end{equation}
Substituting Eq.~\eqref{eq:scale_factor_relation} and Eq.~\eqref{flow_parm_rel} into Eq.~\eqref{eq_z'}, the first derivative simplifies to:
\begin{align}
z'
=
\sqrt{2}\,a^2H\sqrt{\epsilon_1}
\left(1+\frac{\epsilon_2}{2}\right)\label{eq:z_prime_final}.
\end{align}
To prepare for the evaluation of the second conformal time derivative, we note the differentiation of the background tracking equations yields:
\begin{equation}
(aH)'=(aH)^2(1-\epsilon_1)\,,\,\,\,\,\left(a^2H\right)'=a^3H^2(2-\epsilon_1)\,,\,\,\,\,\left(\sqrt{\epsilon_1}\right)'=
\frac{aH\sqrt{\epsilon_1}\epsilon_2}{2}\,.
\label{eq:background_derivs}
\end{equation}
We will use these above relations to compute the second conformal time derivative $z''$. Differentiating the expression in Eq.~\eqref{eq:z_prime_final} with respect to the conformal time:
\begin{align}
z''
&=
\sqrt{2}\left(a^2H\sqrt{\epsilon_1}
\left(1+\frac{\epsilon_2}{2}\right)\right)'
\nonumber\\
&=
\sqrt{2}\left(a^2H\right)'
\sqrt{\epsilon_1}
\left(1+\frac{\epsilon_2}{2}\right)
+
\sqrt{2}\,a^2H
\left(\sqrt{\epsilon_1}\right)'
\left(1+\frac{\epsilon_2}{2}\right)
\nonumber\\
&\quad
+
\sqrt{2}\,a^2H\sqrt{\epsilon_1}
\left(\frac{\epsilon_2'}{2}\right)\label{eq:z_double_prime_expanded}\,.
\end{align}
Substituting the previous results in Eq.~\eqref{eq:z_double_prime_expanded}, using the relation \eqref{flow_parm_rel}  we obtain:
\begin{align}
z''
&=
\sqrt{2}\,a^3H^2(2-\epsilon_1)
\sqrt{\epsilon_1}
\left(1+\frac{\epsilon_2}{2}\right)
+
\sqrt{2}\,a^3H^2
\frac{\sqrt{\epsilon_1}\epsilon_2}{2}
\left(1+\frac{\epsilon_2}{2}\right)
\nonumber\\
&\quad
+
\sqrt{2}\,a^3H^2\sqrt{\epsilon_1}
\frac{\epsilon_2\epsilon_3}{2}\nonumber\\
&=
\sqrt{2}\,a^3H^2\sqrt{\epsilon_1}
\left[
2-\epsilon_1
+\frac{3}{2}\epsilon_2
-\frac12\epsilon_1\epsilon_2
+\frac14\epsilon_2^2
+\frac12\epsilon_2\epsilon_3
\right].
\end{align}
Therefore, we finally obtain
\begin{equation}
\frac{z''}{z}
=
(aH)^2
\left[
2-\epsilon_1
+\frac{3}{2}\epsilon_2
-\frac12\epsilon_1\epsilon_2
+\frac14\epsilon_2^2
+\frac12\epsilon_2\epsilon_3
\right]\label{z"/z}.
\end{equation}
We will use this general expression for our particular case of CRI after some more manipulations with the Hubble-flow parameters.


Using the first relationship in Eq.~(\ref{eq:background_derivs}) we have:
\begin{equation}
\left(\frac{1}{aH}\right)'
=
-(1-\epsilon_1)\,.
\label{intm1}
\end{equation}
Using the constant-roll condition in Eq.~(\ref{ep2_ep1_rel}) and $\epsilon_2=\epsilon_1^\prime/(aH\epsilon_1)$, we have: $\epsilon_1^\prime=aH\epsilon_1(2\epsilon_1 - 6\beta)$.
At zeroth order $aH \simeq -1/\tau$ and consequently $\epsilon_1^\prime = (6\beta/\tau)\epsilon_1 -(2\epsilon_1^2/\tau)$. Here, both the terms contribute as for small values of $\beta$ one may not be able to neglect $\epsilon_1$ with respect to $\beta$. 
In the present case we can write $1/aH$ up to the second order as:
\begin{eqnarray}
    \frac{1}{aH}
    =
    -\tau\left(1-c\epsilon_1+d\epsilon_1^2\right)
    +{\cal O}(\epsilon_1^3).
\end{eqnarray}
Differentiating the ansatz for $1/(aH)$ gives
\[
    \left(\frac{1}{aH}\right)'
    =
    -1
    +c\epsilon_1
    -d\epsilon_1^2
    +c\tau\epsilon_1'
    -2d\tau\epsilon_1\epsilon_1'
    +{\cal O}(\epsilon_1^3).
\]
Substituting the expression for $\tau\epsilon_1'$ into the above equation,
we find
\[
    \left(\frac{1}{aH}\right)'
    =
    -1
    +c(1+6\beta)\epsilon_1
    +
    \left[
        c(6\beta c-2)-d(1+12\beta)
    \right]\epsilon_1^2
    +{\cal O}(\epsilon_1^3).
\]
The exact identity in Eq.~(\ref{intm1})
then requires
\[
    c(1+6\beta)=1,
    \qquad
    c(6\beta c-2)-d(1+12\beta)=0.
\]
Therefore,
\begin{eqnarray}
    c=\frac{1}{1+6\beta}\,,\,\,\,\,\,\,
    d =
    -\frac{2(1+3\beta)}
    {(1+6\beta)^2(1+12\beta)}.
\end{eqnarray}
Using the value of $c$, we write:
\begin{equation}
\frac{1}{aH}
=
-\tau
\left(
1-\frac{\epsilon_1}{1+6\beta}
\right)
+\mathcal{O}(\epsilon_1^2),
\end{equation}
or equivalently,
\begin{equation}
aH
=
-\frac{1}{\tau}
\left(
1+\frac{\epsilon_1}{1+6\beta}
\right)
+\mathcal{O}(\epsilon_1^2).
\end{equation}
Substituting $\epsilon_2=2\epsilon_1-6\beta$, and  $\epsilon_3=2\epsilon_1$, into Eq.~(\ref{z"/z}) for $z''/z$, we obtain
\begin{align}
\frac{z''}{z}
&\simeq
\frac{1}{\tau^2}
\left(
1+\frac{2\epsilon_1}{1+6\beta}
\right)
\Big(
2-9\beta+9\beta^2
+(2
-9\beta)\epsilon_1
\Big)
\nonumber\\
&=
\frac{1}{\tau^2}
\left[
2-9\beta+9\beta^2
-
\frac{(-6+15\beta+36\beta^2)}{1+6\beta}\epsilon_1
\right].
\end{align}
Comparing the above equation with
\begin{equation}
\frac{z''}{z}
=
\frac{\nu^2-\frac14}{\tau^2}\,,
\end{equation}
we obtain
\begin{equation}
\nu^2
=
\frac94
-9\beta+9\beta^2
-
\frac{(-6+15\beta+36\beta^2)}{1+6\beta}\epsilon_1.
\end{equation}
The parameter \(\nu\) denotes the order of the Hankel function appearing in the solution of the Mukhanov–Sasaki equation. It is fixed by comparing the background-dependent quantity \(z''/z\) with \((\nu^2-1/4)/\tau^2\), as done above.

Expanding $\nu$ to first order in $\epsilon_1$, we find
\begin{equation}
\nu
\simeq
\frac{3}{2}
-3\beta
+\left(2-13\beta+64\beta^2\right)\epsilon_1\,.
\end{equation}
The Mukhanov--Sasaki equation then takes the form
\begin{equation}
v_k''+
\left(
k^2-\frac{\nu^2-\frac14}{\tau^2}
\right)v_k=0,
\end{equation}
whose Bunch--Davies solution is
\begin{equation}
v_k
=
\frac{\sqrt{-\pi\tau}}{2}
H_\nu^{(1)}(-k\tau).
\end{equation}
On super-Hubble scales ($-k\tau\ll1$), the scalar power spectrum is given by
\begin{equation}
\mathcal P_{\mathcal R}(k)
=
\frac{2^{\,2\nu-3}}{2M^2_{\rm Pl}\epsilon_1}
\left[
\frac{\Gamma(\nu)}
{\Gamma(3/2)}
\right]^2
\left(
\frac{H}{2\pi}
\right)^2
\left(\frac{k}{aH}\right)^{3-2\nu}.
\end{equation}
Here $\Gamma(\nu)$ is the Gamma function. Consequently, the scalar spectral index is
\begin{equation}
n_s-1
=
3-2\nu.
\end{equation}
Substituting the first-order expression for $\nu$, we obtain
\begin{equation}
n_s-1
=
6\beta
+\left(-4+26\beta-128\beta^2\right)\epsilon_1
\label{eq:ns}.
\end{equation}
This is the expression for the spectral index in CRI. 

\section{Cosmological Parameter Estimation}
\label{sec:data_analysis}

Since the scalar spectral index is defined at the epoch when a given mode exits the Hubble horizon, the slow-roll parameter \(\epsilon_1\), appearing in Eq.~(\ref{eq:ns}), should be evaluated at the corresponding horizon-crossing time \(t_*\), determined by:
\[
k=a(t_*)H(t_*).
\]
In the following, we 
examine the resulting variation of the spectral index, $n_s$, with respect to the constant-roll parameter, \(\beta\). Substituting the expression for $\epsilon_{1}$ from Eq.~\eqref{ep1_@_p.s} into  Eq.~\eqref{eq:ns} we obtain:
\begin{equation}
    n_s=1+6\beta+\left(-4+26\beta-128\beta^2\right)\frac{3\beta}{1-(1-3\beta)e^{-6\beta \tilde N}}\label{eq:ns_N}.
\end{equation}
It shows that in general $n_s=n_s(\beta, \tilde{N})$ is a transcendental function of both $\beta$ and $\tilde{N}$. Since Eq.~(4.1) defines a non-monotonic transcendental map from $\beta$ to $n_s$, the inference of $\beta$ from a measured scalar tilt is not unique. Consequently, the posterior distribution of $\beta$ can develop multiple regions of support, corresponding to different constant-roll branches that
produce nearly the same value of $n_s$. Moreover, the above equation shows that in CRI the spectral index $n_s$ is sensitive to the constant-roll parameter and not to the other factors which may be present inside the inflaton potential $V(\phi)$. Consequently, observational bounds on $n_s$ can only fix probable values of $\beta$ but will not be directly constraining the potential parameters.
\begin{figure}[t]
    \centering
    \includegraphics[width=\columnwidth]{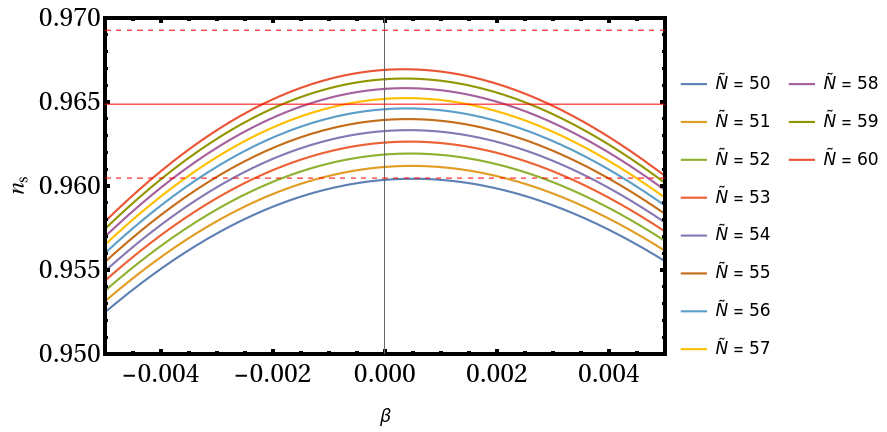}
    \caption{
Scalar spectral index $n_s$ as a function of the constant-roll parameter $\beta$ for $\tilde N=50$--$60$. The solid horizontal line denotes the Planck 2018 mean value of the scalar spectral index, while the dashed horizontal lines indicate the corresponding $68\%$ confidence interval obtained from the Planck 2018 TT, TE, EE+lowE data. The intersections of the theoretical curves with the observational bounds constrain the allowed values of $\tilde N$ for a range of $\beta$.
}
    \label{fig:ns_beta}
\end{figure}
From Fig.~\ref{fig:ns_beta}, we can observe that the theoretical prediction for $\tilde N=50$ lies at the edge of the $68\%$ confidence interval ($1\sigma$) of the Planck 2018 TT, TE, EE+lowE measurement of the scalar spectral index  \cite{Aghanim2020}. This suggests that Eq.~\eqref{eq:ns_N} imposes a lower bound on the number of remaining e-folds required for consistency with the Planck 2018 observations. For the present model, this lower bound is approximately $\tilde{N}\gtrsim 50$.

Before expressing $\epsilon_{1*}$ in terms of $\beta$ and $\tilde N$,
the spectral-index relation is approximately quadratic in $\beta$ for a fixed
value of $\epsilon_1$, as shown in Eq.~(\ref{eq:ns}). However, after substituting the constant-roll evolution of $\epsilon_{1*}$, Eq.~(4.1) becomes a nonlinear transcendental relation because $\beta$ appears both algebraically and in the exponential factor $\exp(-6\beta\widetilde N)$.
To determine the theoretical predictions of the model, we work with the quadratic approximation of the problem and use Eq.~\eqref{eq:ns} for $n_s = 0.9649$. For a given number of remaining $e$-folds $\tilde{N}$, this equation admits multiple distinct solutions corresponding to the negative and positive branches of $\beta$. The two-branch structure arises because $n_s(\beta)$ is non-monotonic near $\beta=0$. In our case we have considered $\tilde{N}=60$, when  
\begin{equation}
    \epsilon_{1*}(60)=\frac{3\beta}
{1-(1-3\beta)e^{-360\beta }}.
\end{equation}
The results for $\tilde N= 60$ are $\beta=-0.002133$ and $\beta=0.002860$ as shown in Fig.~\ref{fig:ns_beta}

\subsection{Constraints from Planck 2018}

\begin{figure}[h]
    \centering
    \includegraphics[width=\columnwidth]{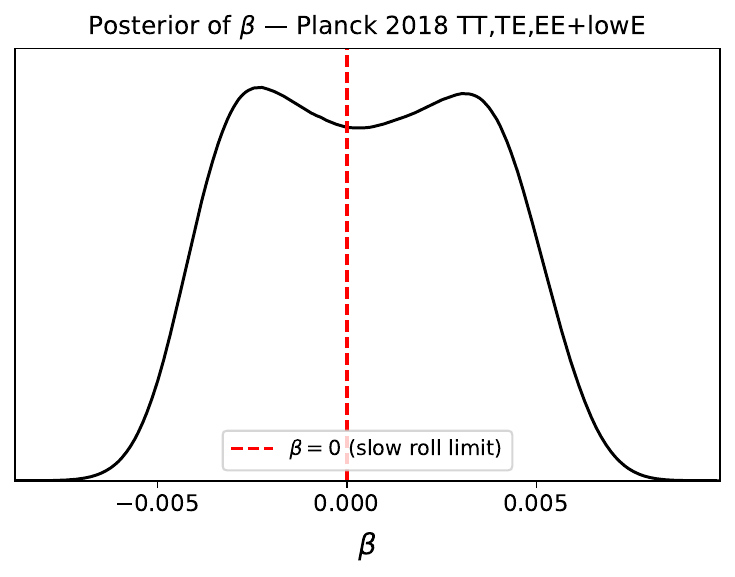}
    \caption{Posterior distribution of the constant-roll parameter $\beta$ 
    from Planck 2018 TT,TE,EE+lowE data. The red dashed line marks 
    $\beta = 0$, corresponding to the slow-roll limit. The bimodal structure 
    reflects the two analytical branches at $\beta_1 \approx -0.002$ and 
    $\beta_2 \approx +0.003$. The two branches cannot be distinguished by 
    current Planck data.}
    \label{fig:beta_posterior}
\end{figure}

\begin{figure}[!t]
    \centering
    \includegraphics[width=\textwidth]{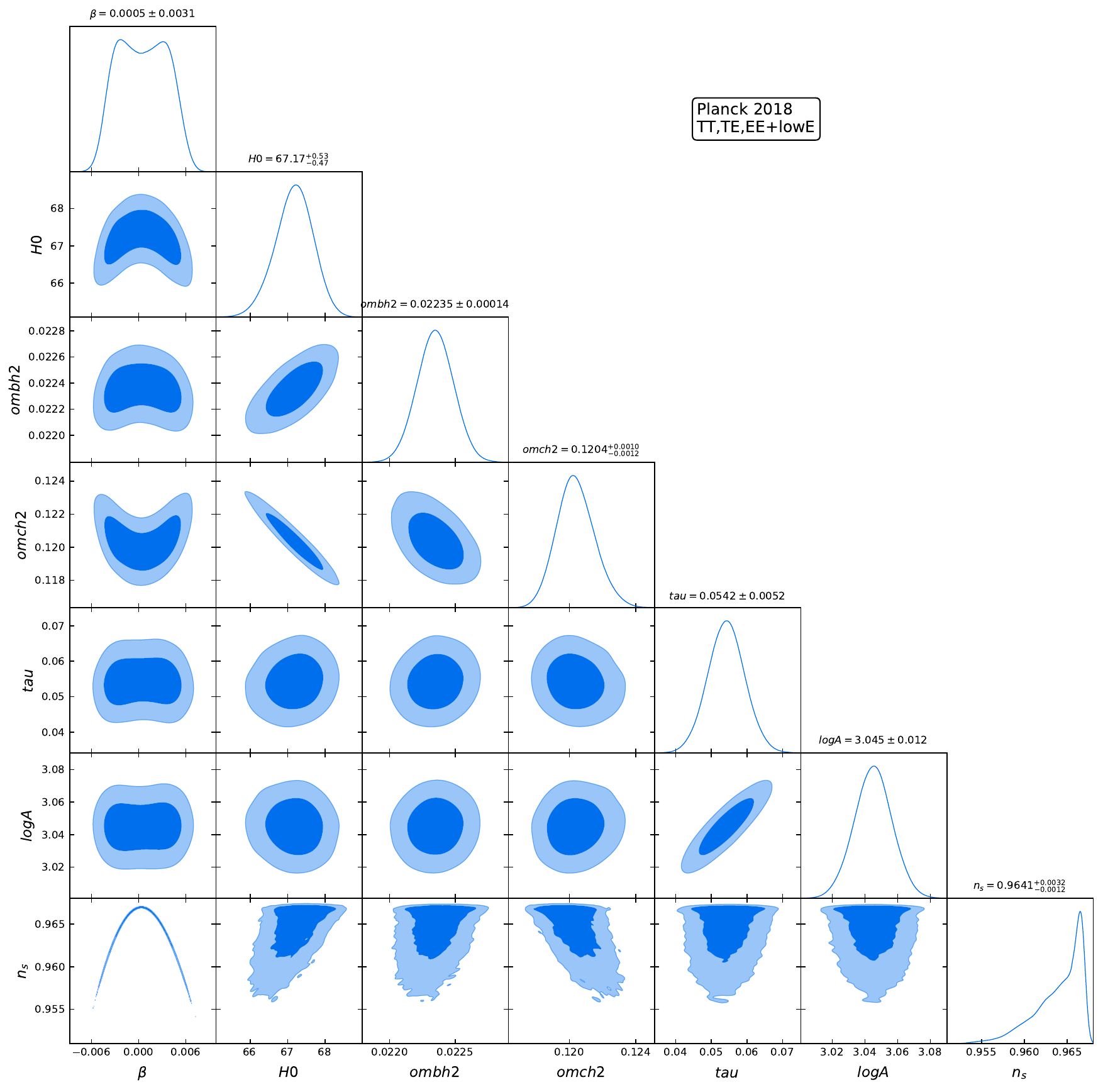}
    \caption{Joint posterior distributions of the constant-roll parameter 
    $\beta$ and the standard $\Lambda$CDM parameters from Planck 2018 
    TT,TE,EE+lowE data. The dark and light shaded regions correspond to 
    the 68\% and 95\% credible intervals respectively. The banana-shaped 
    contours involving $\beta$ reflect the bimodal posterior, with each 
    branch preferring slightly different values of the cosmological 
    parameters. All standard parameters are consistent with the Planck 2018 
    baseline $\Lambda$CDM results.}
    \label{fig:triangle}
\end{figure}

In order to constrain the constant-roll parameter \(\beta\), we performed a Markov Chain Monte Carlo (MCMC) analysis using the \texttt{Cobaya} framework. The analysis employed the Planck 2018 temperature and polarization likelihoods, namely the low-multipole temperature likelihood \texttt{planck\_2018\_lowl.TT}, the low-multipole polarization likelihood \texttt{planck\_2018\_lowl.EE}, and the high-multipole Plik-lite likelihood \texttt{planck\_2018\_highl\_plik.TTTEEE\_lite}. Five independent chains were generated and a burn-in fraction of \(30\%\) was discarded before combining the samples. Five independent chains were generated and a burn-in fraction of \(30\%\) was removed before combining the samples. With $\tilde{N}= 60$ $e$-folds before the end of inflation in Eq.~\eqref{eq:ns_N}, the posterior 
distribution of $\beta$ is shown in Fig.~\ref{fig:beta_posterior}, and the 
joint constraints on all parameters are shown in Fig.~\ref{fig:triangle}.

The posterior distribution of $\beta$ is bimodal, with two peaks 
corresponding to the two analytical branches:
\begin{equation}
\beta_1 \approx -0.00213 \quad \text{(Branch 1, } \beta < 0\text{)},
\qquad
\beta_2 \approx +0.00286 \quad \text{(Branch 2, } \beta > 0\text{)}.
\end{equation}
However, current Planck data cannot distinguish between the two branches, 
since the difference in $n_s$ between them ($\Delta n_s \sim 0.002$) is 
smaller than the Planck measurement uncertainty ($\sigma(n_s) \sim 0.004$). 
The combined constraint on $\beta$ is
\begin{equation}
\beta = 0.0005 \pm 0.0031 \quad (68\%\ \mathrm{CL}),
\end{equation}
\begin{equation}
-0.0047 < \beta < 0.0059 \quad (95\%\ \mathrm{CL}),
\end{equation}
or equivalently $|\beta| < 0.006$ at $95\%$ CL, consistent with the 
slow-roll limit $\beta = 0$.

The constraints on the standard $\Lambda$CDM parameters are consistent 
with the Planck 2018 baseline results:
\begin{equation}
H_0 = 67.17 \pm 0.51\ \mathrm{km\,s^{-1}\,Mpc^{-1}}, \quad
\Omega_b h^2 = 0.02235 \pm 0.00014, \quad
\Omega_c h^2 = 0.1204 \pm 0.0012,
\end{equation}
\begin{equation}
\tau = 0.0542 \pm 0.0052, \quad
n_s = 0.9641^{+0.0032}_{-0.0012}.
\end{equation}
The spectral index is consistent with the Planck 2018 baseline value of 
$n_s = 0.9649 \pm 0.0042$ \cite{Aghanim2020}. The model predicts a hard 
upper limit of $n_s \leq 0.9669$, corresponding to the slow-roll limit 
$\beta = 0$.

The bimodal structure of the posterior distribution of $\beta$, clearly 
visible in Fig.~\ref{fig:beta_posterior}, reflects the existence of two 
distinct analytical branches. However, neither the one-dimensional posterior 
nor the joint two-dimensional contours in Fig.~\ref{fig:triangle} show any 
statistically significant preference for either branch. The posterior 
probability between the two peaks does not vanish, indicating that the 
data cannot rule out any value of $\beta$ in the range 
$-0.005 \lesssim \beta \lesssim 0.006$. Consequently, current Planck 2018 
data cannot differentiate between the two branches, and therefore cannot 
determine which of the two corresponding inflationary potentials is 
observationally favoured. We should keep in mind that the two peek arises because the Eq.~\eqref{eq:ns_N} is insensitive to the potential. Both the potential can give rise to the same $\epsilon_1(N)$. So when we look at Fig.~\ref{fig:beta_posterior} we see that the positive $\beta$ peak is for one potential and the negative $\beta$ beta peak is for a different potential. This degeneracy arises because the $n_s$-$\beta$ 
relation has a maximum at $\beta = 0$ and is extremely flat in the vicinity 
of both branches, with the difference in $n_s$ between them being 
$\Delta n_s \sim 0.002$, which is smaller than the current Planck 
measurement uncertainty of $\sigma(n_s) \sim 0.004$. Breaking this 
degeneracy will require different techniques or future CMB experiments such as CMB-S4 
\cite{abazajian2019cmbs4sciencecasereference} and LiteBIRD \cite{Matsumura_2014}, which are expected to achieve 
$\sigma(n_s) \sim 0.002$, sufficient to resolve the two branches.

\subsection{Planck 2018 + ACT DR6 data}

We further constrain the constant-roll parameter $\beta$ using the 
combined Planck 2018 + ACT DR6 dataset. Specifically, we use the 
Planck 2018 low-$\ell$ TT and EE likelihoods, the Planck 2018 
high-$\ell$ \texttt{Plik} lite TT,TE,EE likelihood, and the ACT DR6 
CMB-only likelihood \cite{ACTCosmo2025}. Unlike the ACT-only analysis, we 
sample $\tau$ with a Gaussian prior from the Planck low-$\ell$ EE 
likelihood, and we sample $100\theta_\mathrm{MC}$ instead of $H_0$ 
following the standard Planck convention.

\begin{figure}[H]
    \centering
    \includegraphics[width=0.6\textwidth]{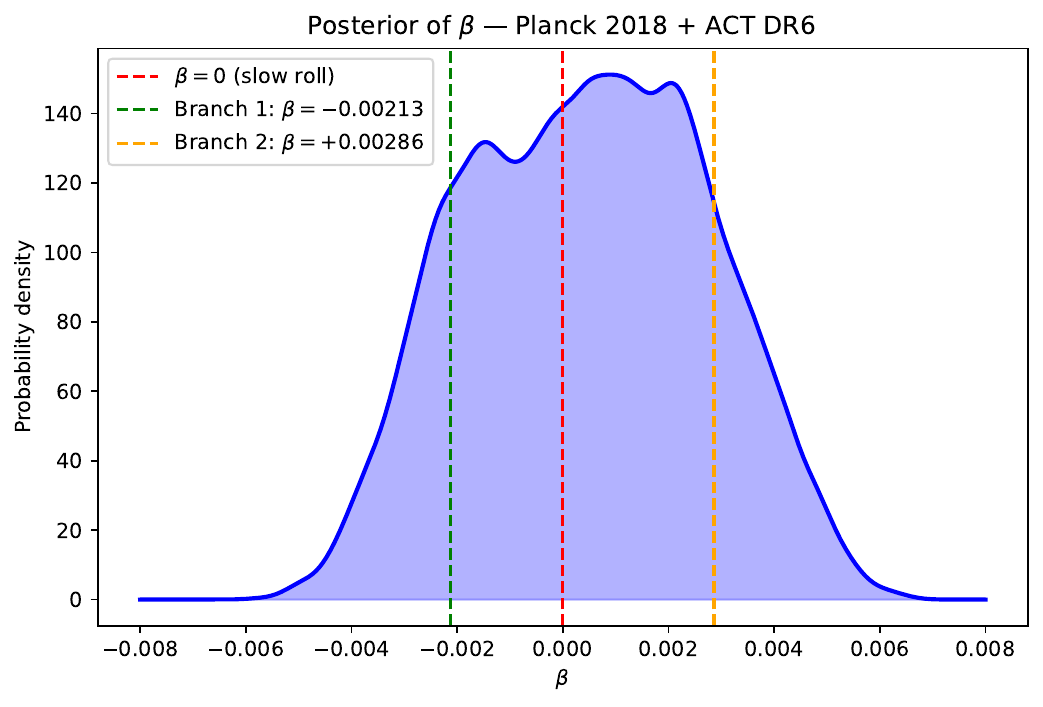}
    \caption{Posterior distribution of the constant-roll parameter 
    $\beta$ from Planck 2018 + ACT DR6 data, estimated using a 
    Gaussian kernel density estimator with bandwidth $h = 0.12$. 
    The red, green and orange dashed lines mark the slow-roll limit 
    $\beta = 0$ and the two analytical branch solutions 
    $\beta_1 \approx -0.002$ and $\beta_2 \approx +0.003$ 
    respectively. The trimodal structure reflects the partial 
    resolution of the two branches by the combined dataset.}
    \label{fig:beta_planck_act}
\end{figure}

\begin{figure}[H]
    \centering
    \includegraphics[width=\textwidth]{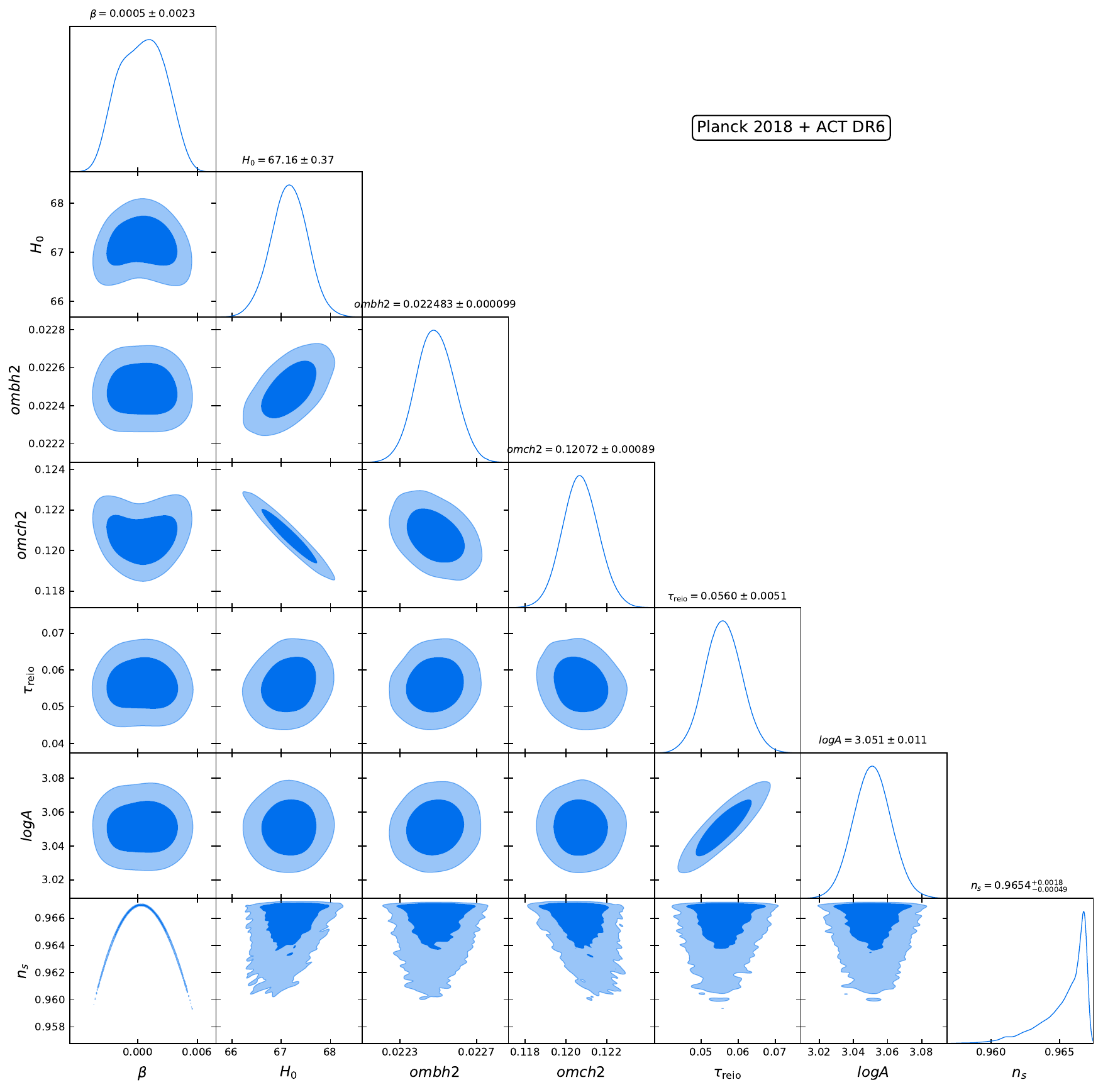}
    \caption{Joint posterior distributions of the constant-roll 
    parameter $\beta$ and the standard $\Lambda$CDM parameters from 
    Planck 2018 + ACT DR6 data. The dark and light shaded regions 
    correspond to the $68\%$ and $95\%$ credible intervals 
    respectively. All cosmological parameters are consistent with the 
    Planck 2018 baseline $\Lambda$CDM results. The crescent-shaped 
    contours in the $\beta$-$H_0$ and $\beta$-$\Omega_c h^2$ panels 
    reflect the trimodal structure of the $\beta$ posterior.}
    \label{fig:triangle_planck_act}
\end{figure}

The posterior distribution of $\beta$ from Planck 2018 + ACT DR6 is 
shown in Fig.~\ref{fig:beta_planck_act} and the joint constraints are 
shown in Fig.~\ref{fig:triangle_planck_act}. The combined constraint is
\begin{equation}
\beta = 0.0005 \pm 0.0023 \quad (68\%\ \mathrm{CL}),
\end{equation}
\begin{equation}
-0.0037 < \beta < 0.0046 \quad (95\%\ \mathrm{CL}),
\end{equation}
or equivalently $|\beta| < 0.0046$ at $95\%$ CL. This is the tightest 
constraint on $\beta$ obtained in this work, improving upon the 
Planck-only result by approximately $25\%$.

The posterior distribution of $\beta$ reveals a trimodal structure, 
with peaks near the two analytical branches at $\beta_1 \approx -0.002$ 
and $\beta_2 \approx +0.003$, and near the slow-roll limit $\beta = 0$. 
This structure reflects the partial resolution of the two branches by 
the combined Planck + ACT dataset, which provides tighter constraints 
on $n_s$ than either dataset alone. The slow-roll limit $\beta = 0$ 
remains consistent with the data at the $68\%$ CL.

The constraints on the standard $\Lambda$CDM parameters are
\begin{align}
H_0 &= 67.16 \pm 0.37\ \mathrm{km\,s^{-1}\,Mpc^{-1}}, \quad
\Omega_b h^2 = 0.02248 \pm 0.00010, \\
\Omega_c h^2 &= 0.1207 \pm 0.0009, \quad
\tau = 0.0560 \pm 0.0051, \quad
n_s = 0.9654^{+0.0018}_{-0.0005}.
\end{align}
\begin{equation}
\tau = 0.0560 \pm 0.0051, \quad
n_s = 0.9654^{+0.0018}_{-0.0005}.
\end{equation}
All parameters are consistent with the Planck 2018 baseline 
$\Lambda$CDM results \cite{Aghanim2020}. The inferred value of $H_0$ 
is in excellent agreement with the Planck-only result, confirming that 
the addition of ACT DR6 data does not introduce any tension in the 
Hubble constant. The spectral index $n_s$ is significantly better 
constrained than in the Planck-only analysis ($\sigma(n_s) = 0.0018$ 
vs $0.0028$), reflecting the additional constraining power of ACT DR6 
on small angular scales.

The three-peak structure in the Planck 2018+ACT DR6 posterior may be
understood as a consequence of the non-monotonic relation between
$\beta$ and $n_s$ in Eq.~(4.1). The two side peaks are naturally associated
with the negative- and positive-$\beta$ constant-roll branches, while the
central peak reflects the continued support for the nearly slow-roll
spectral point $\beta=0$. Since the posterior shape can depend on the
kernel density estimate, sampling, priors, and correlations with the
standard cosmological parameters, this trimodal structure should be
interpreted as suggestive rather than as decisive evidence for three
separate physical branches.

The addition of ACT DR6 reduces the overall allowed range of $\beta$, but
this need not make the posterior single-peaked. Since the map
$n_s(\beta,\widetilde N)$ is non-monotonic, several nearby values of
$\beta$ can produce very similar scalar tilts. The two side peaks may be
associated with the negative- and positive-$\beta$ analytical branches,
while the central peak reflects the continued support for the nearly
slow-roll spectral point $\beta=0$. Thus the Planck+ACT posterior is not only more
concentrated in width, but more structured internally. The detailed
trimodal appearance should nevertheless be interpreted cautiously.

\subsection{Comparison of Goodness of Fit and Model Preference }
\begin{table*}
    \centering
    \caption{Table showing marginalized best-fit cosmological parameter values for constant-roll and slow-roll inflation models based upon Planck and ACT observations.}
    \label{tab:params}
    \renewcommand{\arraystretch}{1.3}

    \resizebox{\textwidth}{!}{%
    \begin{tabular}{c c c c c}
        \hline
        \hline
        Cosmological
        & Constant roll + Planck
        & Constant roll + Planck + ACT
        & Slow roll + Planck
        & Slow roll + Planck + ACT \\
        Parameters
        & 
        & 
        & 
        & \\
        \hline

        $\beta$
        & $0.000494_{-0.002949}^{+0.003940}$
        & $0.000452_{-0.002053}^{+0.002836}$
        & $--$
        & $--$ \\

        $H_0$
        & $67.168_{-0.465}^{+0.534}$
        & $67.162_{-0.364}^{+0.368}$
        & $67.291_{-0.610}^{+0.589}$
        & $67.314_{-0.434}^{+0.434}$ \\

        $\Omega_bh^2$
        & $0.022349_{-0.000139}^{+0.000139}$
        & $0.022483_{-0.000098}^{+0.000100}$
        & $0.022371_{-0.000141}^{+0.000141}$
        & $0.022490_{-0.000096}^{+0.000098}$ \\

        $\Omega_ch^2$
        & $0.120391_{-0.001217}^{+0.001036}$
        & $0.120715_{-0.000881}^{+0.000881}$
        & $0.120132_{-0.001354}^{+0.001384}$
        & $0.120336_{-0.001034}^{+0.001049}$ \\

        $\tau$
        & $0.054194_{-0.005117}^{+0.005120}$
        & $0.055987_{-0.005014}^{+0.004975}$
        & $0.054456_{-0.005116}^{+0.005151}$
        & $0.056509_{-0.005149}^{+0.005132}$ \\

        $n_s$
        & $0.964079_{-0.001196}^{+0.003177}$
        & $0.965379_{-0.000493}^{+0.001794}$
        & $0.965254_{-0.004193}^{+0.004155}$
        & $0.967187_{-0.003471}^{+0.003482}$ \\

                $A_s$
        & $(2.1011_{-0.0248}^{+0.0248})\times10^{-9}$
        & $(2.1138_{-0.0238}^{+0.0236})\times10^{-9}$
        & $(2.1005_{-0.0245}^{+0.0239})\times10^{-9}$
        & $(2.1136_{-0.0237}^{+0.0239})\times10^{-9}$ \\

        \hline
    \end{tabular}%
    }
\end{table*}

In table~\ref{tab:params} we show the best fit values of basic cosmological parameters from the individual marginalized likelihood functions along with $1\sigma$ upper and lower confidence interval obtained by using observations from Planck and ACT telescopes. While constraining CRI model of this work we do not sample $n_s$ directly. Instead, since $n_s$ is completely determined by $\beta$ we sample $\beta$ along with $5$ other standard cosmological parameters, namely, $\{H_0, \Omega_bh^2, \Omega_ch^2, \tau, A_s$\}. The value of $n_s$ for each sampled $\beta$ can then be calculated using Eqn.~\ref{eq:ns_N}.  The third column represents the constrains on the standard $6$ parameter LCDM (flat) model, where $n_s$ was also sampled along with rest $5$ parameters.  As we see from this table, values of all common cosmological parameters across the three columns agree very well with each other. It is interesting to note that values of $n_s$ from the CRI model (with Planck + ACT observations) agree remarkably well with its value obtained for slow roll inflationary model. Value of $n_s$ for CRI model with Planck data alone agree with the results of other two columns close to $1\sigma$ error limit. As expected, all cosmological parameters are constrained better when Planck observation was supplemented  by ACT.

Before ending this discussion  it will be interesting to see if our model provides a better fit  to the Planck data compared to the Planck's 6 parameter model. We estimate the best fit values of cosmological parameters along with $\chi^2$ value defined by negative twice of log joint likelihood function at the best fit cosmological parameters. We compare  $\chi^2$  obtained from CRI model along with the corresponding value for slow roll inflation model using Plank observations. We find that the best-fit $\chi^2$ for our constant-roll model is $\chi^2_{\rm min} = 1002.17$, compared to $\chi^2_{\rm min} = 1003.03$ for the standard slow roll inflation model while using the same data set. The improvement $\Delta\chi^2 = -0.86$ indicates that our model unambiguously provides at least as equivalent (or marginally better) fit to latest CMB observations as the slow roll model of inflation~\footnote{Since both the constant roll and slow roll models have the same number of sampled parameters information based criterion to assess the goodness of fit is not necessary.}. Following the same procedure, we also estimated the best-fit $\chi^2$ values using the combined Planck + ACT dataset. We find $\chi^2_{\rm min} = 1160.59$ for the constant-roll model, compared to $\chi^2_{\rm min} = 1160.47$ for the standard slow-roll model, giving $\Delta\chi^2 = +0.12$. This indicates that the two models fit the combined Planck + ACT data essentially equally well, in contrast to the Planck-alone case where the constant-roll model howed a marginally better fit ($\Delta\chi^2 = -0.86$).
\section{Estimation of parameters in the inflaton potential: Values of $M$ and $C$}
\label{param}

Using MCMC, we sampled $\beta$ and derived the $n_s$ values in the previous section. It is seen that this process does not yield any information about the parameters $M$ and $C$ appearing in the inflaton potential. Out of these two parameters, it will be seen that when $\beta>0$, $C$ is just a coordinate time origin shifting parameter and does not give rise to any new physics. The parameter $M$ affects the physics of the system. In the present section the mass scale $M$ is not sampled as an independent MCMC parameter. Instead, it is derived a posteriori from the sampled values of $A_s \equiv \mathcal P_{\mathcal R}(k_*)$ and $\beta$ using the scalar-amplitude normalization and the branch-specific background expression for $H(\tilde N)$. Here $k^*$ is the pivot comoving wavenumber at which the observed amplitude $A_s$ is given. Physically, $k_*$ is the mode that leaves the Hubble horizon at $t_*$: $k_*=a(t_*)H(t_*)$. Usually, for Planck analyses, the pivot is taken to be $k_*=0.05\,{\rm Mpc}^{-1}$.

\subsection{Case1: $\beta <0, \ \ \tilde \beta  =- \beta > 0$}

From our previous discussions it is known that:
\begin{equation}
\mathcal P_{\mathcal R}(k_*)  \equiv A_s =\frac{2^{1-n_s}}{2M^2_{Pl}\epsilon^*_1\pi^3}\Gamma^2(\nu) H^2(\tilde{N}_*)\,,
\end{equation}
where we fix $\tilde{N} = 60$, the functional form of $\epsilon^*_1$ is given from Eq.~(\ref{ep1_@_p.s}) and $\nu = 2-n_s/2$. 
The value of the Hubble parameter at the pivot scale can be obtained from the observed scalar-amplitude normalization. 
At horizon crossing, $k_*=a_*H_*$, and therefore
\begin{equation}
    A_s
    =
    \frac{2^{2\nu-3}}{2M_{\rm Pl}^2\epsilon_{1*}}
    \left[
        \frac{\Gamma(\nu)}{\Gamma(3/2)}
    \right]^2
    \left(
        \frac{H_*}{2\pi}
    \right)^2 ,
\end{equation}
where $H_* \equiv H(\tilde{N_*}), \epsilon_{1*}\equiv \epsilon_1(\tilde{N_*})$, and using $\Gamma(3/2)=\frac{\sqrt{\pi}}{2}$, we can rewrite the above equation as
\begin{align}
    A_s
= \frac{2^{2\nu-4}\Gamma^2(\nu)} {\pi^3M_{\rm Pl}^2\epsilon_{1*}} H_*^2 \,.
\end{align}
Therefore,
\begin{equation}
    H_*^2
    =
    A_s\,
    \pi^3M_{\rm Pl}^2\epsilon_{1*}\,
    \frac{2^{n_s}}{\Gamma^2(\nu)} .
\end{equation}
For the negative $\beta$ branch the Hubble scale at the pivot is:
\begin{equation}
    H^2(\tilde N_*)
    =
    A_s\,
    \pi^3M_{\rm Pl}^2
    \left[
    \frac{3\tilde\beta}
    {(1+3\tilde\beta)e^{6\tilde\beta\tilde N_*}-1}
    \right]
    \frac{2^{n_s}}{\Gamma^2\!\left(2-\frac{n_s}{2}\right)}\,.
\end{equation}
For illustration, using the central values
$A_s=2.101\times10^{-9}$, $n_s=0.9641$,
$\tilde N_*=60$, and $\tilde\beta=0.00213$,
one obtains $H^2(\tilde N_*)=5.2466\times10^{27}\,{\rm GeV}^2$.
For the posterior distribution of $M$, however, we do not keep
$A_s$ and $\tilde\beta$ fixed at these central values. Instead,
for each Planck MCMC sample with $\beta<0$, we compute $M$ from the scalar-amplitude
normalization together with the negative-branch background relation.

Using $H(t) = -M\tanh(3\tilde \beta Mt)$ in above equation we obtain
\begin{eqnarray}
    \tilde{N}_* =-\frac{1}{3\tilde \beta} \int_{t^*}^{t_e} \frac{d(\cosh(3\tilde \beta M t)}{\cosh(3\tilde \beta M t)} = -\frac{1}{3\tilde \beta}\ln\biggl[\frac{\cosh(3\tilde \beta M t_e)}{\cosh(3\tilde \beta M t^*)}\biggr]\,,
\end{eqnarray}
where $t^*$ denotes the time at which a cosmological mode of interest with comoving wavenumber $k^*$ becomes super-horizon type during inflation. The above equation gives 
\begin{eqnarray}
    \cosh(3M\tilde \beta t^*) =  \cosh(3M\tilde \beta t_e) \exp(3N_*\tilde \beta)\,.
\end{eqnarray}
We can estimate the time of the end of inflation $t_e$ following
\begin{eqnarray}
t_e = \frac{1}{3M\tilde \beta}\tanh^{-1}\biggl[-\sqrt{\frac{3\tilde \beta}{1+3\tilde \beta}}\biggr]\,,    
\end{eqnarray} 
which can be derived easily using the condition $\epsilon_1 = 1$ at the end of inflation. Using $\tilde \beta = 0.00213$ for the negative $\beta$ (as is the case here) one finds   $3M\tilde \beta t_e = 0.0798525 \equiv K$, where this $K$ is a constant irrespective of the value of $M$, as shown from the above equation, and we have $\cosh(K) = 1.0032$. One can show that:
\begin{eqnarray}
 H(\tilde{N}_*) = M\frac{\sqrt{\cosh^2(3M\tilde \beta t^*)-1}}{\cosh(3M\tilde \beta t^*)}\,, 
\end{eqnarray}
Using the above equation one gets an estimate of $M$ if one knows the values of the other dynamical quantities at the time when the pivot scale corresponding to $k_*$ crossed the Hubble horizon. For the negative-$\beta$ branch for
$\tilde N_* = 60$ the scalar-amplitude normalization gives:
\[
    H_*^2
    =
    A_s\pi^3M_{\rm Pl}^2\epsilon_{1*}
    \frac{2^{n_s}}{\Gamma^2(\nu)},
    \qquad
    \nu=2-\frac{n_s}{2}.
\]
Using the appropriate form of the Hubble-flow parameter, for this branch:
\[
    H_*^2
    =
    M^2
    \frac{(1+3\tilde\beta)e^{6\widetilde\beta\tilde N_*}-1}
    {(1+3\tilde\beta)e^{6\tilde\beta\tilde N_*}},
\]
we obtain
\[
    M^2
    =
    A_s\,
    \frac{6\pi^3M_{\rm Pl}^2\tilde\beta}{2^{1-n_s}\Gamma^2(\nu)}
    \,
    \frac{(1+3\tilde\beta)e^{6\tilde\beta\widetilde N_*}}
    {\left[(1+3\tilde\beta)e^{6\tilde\beta\tilde N_*}-1\right]^2}.
\]
Thus, for each sampled value of $A_s$ and $\beta<0$, the mass scale $M$ is determined after fixing
$\tilde N_*$.

\begin{figure}[t]
    \centering
    \includegraphics[width=0.8\textwidth]{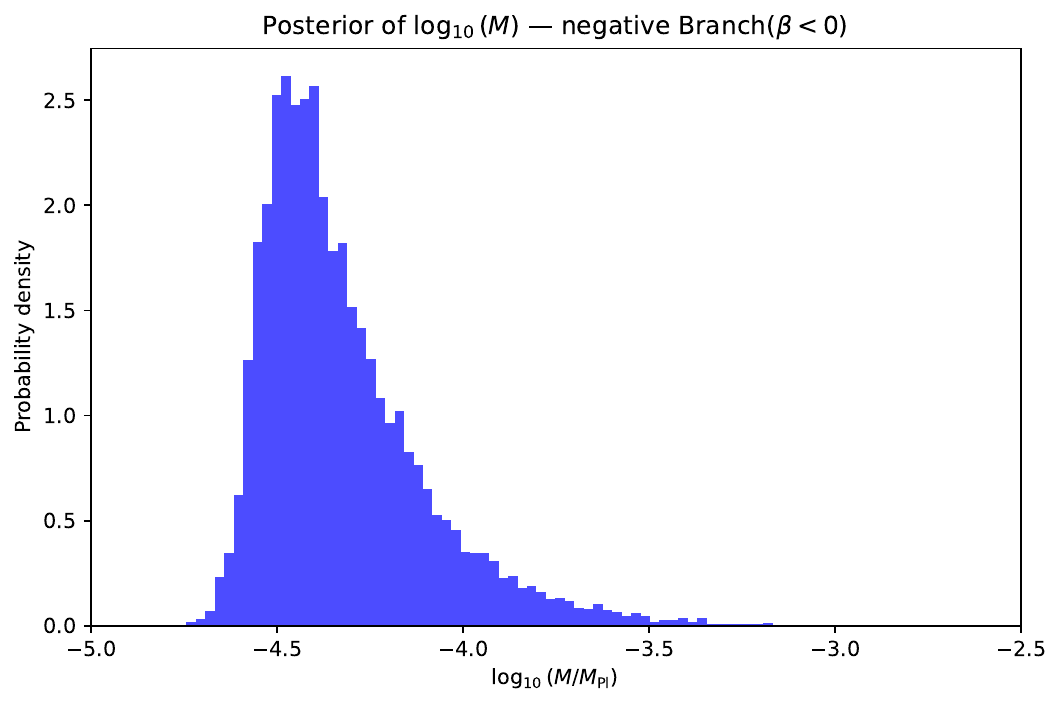}
    \caption{Posterior distribution of $\log_{10}(M/M_{\rm Pl})$ for 
    the negative branch ($\beta < 0$) of the constant-roll inflation 
    model, derived from the Planck 2018 MCMC chains. The distribution 
    is peaked at $\log_{10}(M/M_{\rm Pl}) \approx -4.5$ with a 
    long tail toward larger values arising from samples near 
    $\beta \approx 0$.}
    \label{fig:M_posterior_b1}
\end{figure}

Using the posterior samples of $A_s$ and $\tilde\beta$ from the Planck 2018 
MCMC chains, we compute the posterior distribution of $M$ for all 
samples with $\beta < 0$. The result is shown in 
Fig.~\ref{fig:M_posterior_b1}. The constraint on $M$ is
\begin{equation}
\log_{10}(M/M_{\rm Pl}) = -4.32 \pm 0.23 \quad (68\%\ \mathrm{CL}),
\end{equation}
\begin{equation}
-4.60 < \log_{10}(M/M_{\rm Pl}) < -3.73 \quad (95\%\ \mathrm{CL}),
\end{equation}
corresponding to
\begin{equation}
M = 4.75 \times 10^{-5}\, M_{\rm Pl} \quad (\text{mean}),
\end{equation}
\begin{equation}
2.49 \times 10^{-5}\, M_{\rm Pl} < M < 1.87 \times 10^{-4}\, M_{\rm Pl}
\quad (95\%\ \mathrm{CL}).
\end{equation}
This shows that the negative-$\beta$ branch corresponds to a high-scale
inflationary normalization. Thus the negative branch is not only compatible
with the observed scalar tilt, but also admits a physically reasonable
amplitude normalization for the reconstructed constant-roll potential.

\subsection{Case 2: $\beta > 0$}
 
In this case $H(t) = M\tan(C-3\beta Mt)$. Following the same approach as in the previous Section we obtain the following
\begin{eqnarray}
\cos(C-3\beta M t ^*) = \cos(C-3\beta M t_e ) \exp(-3\tilde{N_*} \beta)\,.
\label{bp1}
\end{eqnarray}
Note the negative exponent in the above equation. For a fixed value of $t_e$,
increasing $\tilde N_*$ corresponds to choosing an earlier horizon-exit time $t_*$, since $\widetilde N_*$ counts 
the number of $e$-folds remaining between $t_*$ and the end of inflation. Therefore $t_*$ decreases as
$\tilde N_*$ increases.
The end of inflation occurs at $t=t_e$, where $\epsilon_1(t_e)=1$, and consequently:
\[
\epsilon_1= -\frac{\dot{H}}{H^2}=\frac{3\beta}{\sin^2 (C-3\beta Mt)}.
\]
Therefore $\sin^2(C-3\beta Mt_e)=3\beta$. Choosing the first-quadrant branch with
$H>0$, we obtain
\[
    \sin(C-3\beta Mt_e)=\sqrt{3\beta},
    \qquad
    \cos(C-3\beta Mt_e)=\sqrt{1-3\beta}.
\]
Hence we also have $\cos(C-3\beta M t_e) = \sqrt{1-3\beta}$.  From Eqn.~(\ref{bp1}) we get 
 \begin{eqnarray}
     \cos(C - 3\beta M t^*) = \sqrt{1 -3\beta} \exp({-3\tilde{N}_*\beta})\,.
 \end{eqnarray}
 Using above we find 
 \begin{eqnarray}
      \sin(C - 3\beta M t^*) = \left(1-(1 -3\beta)
      \exp(-6\tilde N\beta)\right)^{1/2}
 \end{eqnarray}
 Finally, 
 \begin{eqnarray}
     H^2(\tilde N) =M^2\frac{\bigl[1-(1-3\beta)\exp(-6\tilde N\beta)\bigr]}{(1-3\beta)\exp(-6\tilde N\beta) }
 \end{eqnarray}
For the appropriate expression of the Hubble-flow parameter for the positive-$\beta$ branch, we have:
\[
    H_*^2
    =
    M^2
    \frac{
    1-(1-3\beta)e^{-6\beta\tilde N_*}
    }{
    (1-3\beta)e^{-6\beta\tilde N_*}
    } .
\]
Therefore,
\[
    M^2
    =
    A_s\,
    \frac{
    6\pi^3M_{\rm Pl}^2\beta
    }{
    2^{1-n_s}\Gamma^2(\nu)
    }
    \frac{
    (1-3\beta)e^{-6\beta\tilde N_*}
    }{
    \left[1-(1-3\beta)e^{-6\beta\tilde N_*}\right]^2
    } .
\]
In this case if we fix $\tilde{N}_*=60$ then the above relation can give possible range of $M$ when $A_s$ and $\beta$ are  sampled.

\begin{figure}[t]
    \centering
    \includegraphics[width=0.6\textwidth]{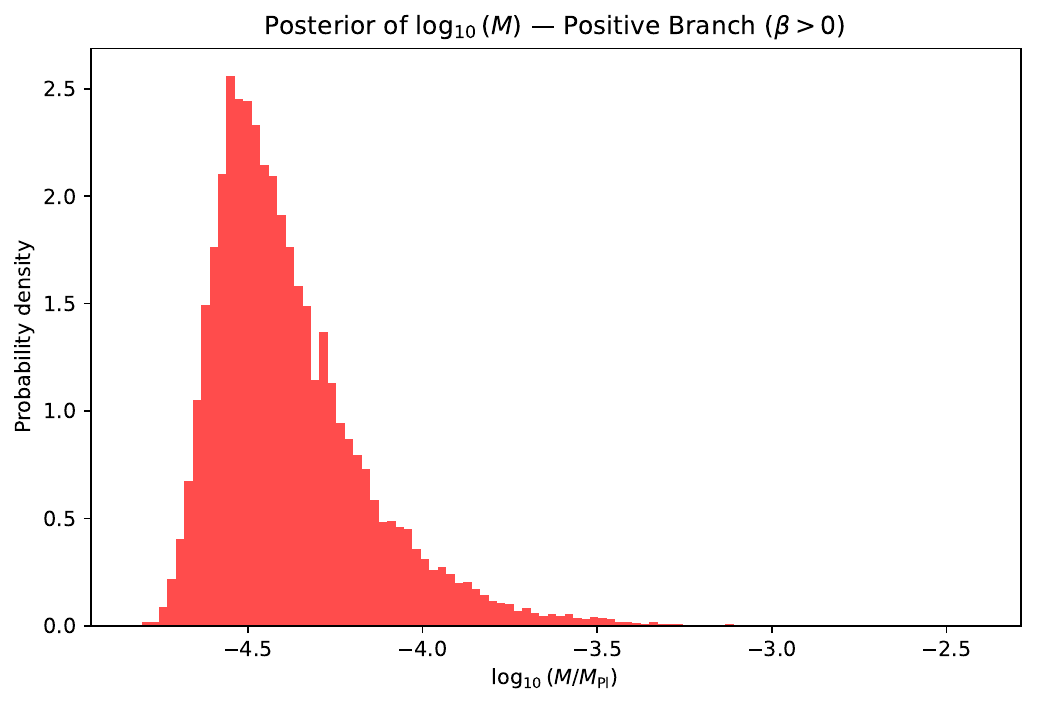}
    \caption{Posterior distribution of $\log_{10}(M/M_{\rm Pl})$ for 
    the positive branch ($\beta > 0$) of the constant-roll inflation 
    model, derived from the Planck 2018 MCMC chains. The distribution 
    is peaked at $\log_{10}(M/M_{\rm Pl}) \approx -4.5$, consistent 
    with the negative branch result, confirming that both branches 
    predict a similar energy scale for the inflationary potential.}
    \label{fig:M_posterior_b2}
\end{figure}

Using the posterior samples of $A_s$ and $\beta$ from the Planck 2018 
MCMC chains, we compute the posterior distribution of $M$ for all 
samples with $\beta > 0$. The result is shown in 
Fig.~\ref{fig:M_posterior_b2}. The constraint on $M$ is
\begin{equation}
\log_{10}(M/M_{\rm Pl}) = -4.38 \pm 0.23 \quad (68\%\ \mathrm{CL}),
\end{equation}
\begin{equation}
-4.67 < \log_{10}(M/M_{\rm Pl}) < -3.80 \quad (95\%\ \mathrm{CL}),
\end{equation}
corresponding to
\begin{equation}
M = 4.20 \times 10^{-5}\, M_{\rm Pl} \quad (\text{mean}),
\end{equation}
\begin{equation}
2.14 \times 10^{-5}\, M_{\rm Pl} < M < 1.58 \times 10^{-4}\, M_{\rm Pl}
\quad (95\%\ \mathrm{CL}).
\end{equation}
Thus the positive-$\beta$ branch is not only
compatible with the observed scalar tilt, but also yields a physically
reasonable amplitude normalization for the reconstructed potential.

The constant $C$ appearing in $H(t)=M\tan(C-3\beta Mt)$ is not an
independent physical parameter. It fixes the origin of time, or equivalently
the phase of the tangent branch. In the present case
one obtains
\[
    \epsilon_1
    =
    -\frac{\dot H}{H^2}
    =
    \frac{3\beta}{\sin^2(C-3\beta Mt)}.
\]
The end of inflation is defined by $\epsilon_1(t_e)=1$, and therefore: $\sin^2(C-3\beta Mt_2)=3\beta$.
Choosing the first-quadrant branch, for which $H>0$, gives: $C=3\beta Mt_e+\sin^{-1}\sqrt{3\beta}$.
Since the time origin is arbitrary, one may set $t_e=0$, in which case $C=\sin^{-1}\sqrt{3\beta}$. Thus $C$ is fixed by the choice of branch and time origin, and it drops out of the final expression for $H^2(\tilde N)$. With fixed \(M\) and fixed \(\beta>0\), different choices of \(C\) will change the coordinate value of \(t_e\), but this is only a coordinate/time-origin shift, not a new physical solution.
\section{The slow-roll limit of constant-roll dynamics}
\label{sll}

It is useful to distinguish two different senses in which the constant-roll
model approaches the slow-roll regime. These two limits coincide at the level
of the scalar spectral index, but they need not coincide at the level of the
reconstructed background dynamics.

In standard single-field slow-roll inflation, the scalar curvature power
spectrum at first order is evaluated at Hubble crossing, $k=aH$, and is given
by
\begin{equation}
    \mathcal P_{\mathcal R}(k)
    \simeq
    \left.
    \frac{H^2}{8\pi^2 M_{\rm Pl}^2\epsilon_1}
    \right|_{k=aH},
\end{equation}
where $\epsilon_1=-\dot H/H^2$.
The scalar spectral index is defined by
\begin{equation}
    n_s-1
    =
    \frac{d\ln \mathcal P_{\mathcal R}}{d\ln k}.
\end{equation}
Taking the logarithm of the scalar spectrum, we obtain $\ln \mathcal P_{\mathcal R} = 2\ln H - \ln \epsilon_1 + \text{constant}$. Let $N=\ln a$ denote the number of $e$-folds increasing forward in time. Then
\begin{equation}
    \frac{d\ln H}{dN}
    =
    \frac{1}{H}\frac{dH}{dN}
    =
    \frac{\dot H}{H^2}
    =
    -\epsilon_1.
\end{equation}
Also, the second Hubble-flow parameter is defined as
\begin{equation}
    \epsilon_2
    =
    \frac{d\ln \epsilon_1}{dN}.
\end{equation}
Therefore,
\begin{align}
    \frac{d\ln \mathcal P_{\mathcal R}}{dN} =
    2\frac{d\ln H}{dN} -
    \frac{d\ln \epsilon_1}{dN} = -2\epsilon_1-\epsilon_2 .
\end{align}
Since the spectrum is evaluated at Hubble crossing, $k=aH$, we have, $\ln k = \ln a+\ln H = N+\ln H$.
Differentiating with respect to $N$ gives
\begin{equation}
    \frac{d\ln k}{dN}
    =
    1+\frac{d\ln H}{dN}
    =
    1-\epsilon_1.
\end{equation}
Therefore,
\begin{equation}
    n_s-1
    =
    \frac{d\ln \mathcal P_{\mathcal R}}{d\ln k}
    =
    \frac{d\ln \mathcal P_{\mathcal R}/dN}
         {d\ln k/dN},
\end{equation}
and so
\begin{equation}
    n_s-1
    =
    \frac{-2\epsilon_1-\epsilon_2}{1-\epsilon_1}.
\end{equation}
At first order in slow-roll, terms of order $\epsilon_1^2$ and
$\epsilon_1\epsilon_2$ are neglected, so
\begin{equation}
    n_s-1
    \simeq
    -2\epsilon_1-\epsilon_2.
\label{nsm1s}    
\end{equation}
This is the standard first-order slow-roll result written in terms of the
Hubble-flow parameters.

Using the leading-order slow-roll relations between the Hubble-flow and
potential slow-roll parameters, $\epsilon_1 \simeq \epsilon_V$, and $\epsilon_2 \simeq 4\epsilon_V-2\eta_V$,
where
\begin{equation}
    \epsilon_V
    =
    \frac{M_{\rm Pl}^2}{2}
    \left(
    \frac{V_{,\phi}}{V}
    \right)^2,
    \qquad
    \eta_V
    =
    M_{\rm Pl}^2
    \frac{V_{,\phi\phi}}{V},
\end{equation}
$V(\phi)$  being the potential responsible for slow-roll inflation, we obtain
\begin{align}
    n_s-1
    &\simeq
    -2\epsilon_V
    -
    \left(
    4\epsilon_V-2\eta_V
    \right)=
    2\eta_V-6\epsilon_V\,.
\end{align}
which is the standard relation one obtains in slow-roll inflation. 

From Eq.~(\ref{eq:ns}) it is seen that as $\beta \to 0$ we have $n_s-1 \simeq -4\epsilon_1$. But does it match with the result obtained from Eq.~(\ref{nsm1s})? Apparently they do not as in the slow-roll case we have two Hubble-flow parameters whereas we have only $\epsilon_1$ in our $\beta \to 0$ limit. This limit shows the subtle difference between the $\beta \to 0$ of CRI and slow-roll inflation. One can match the $n_s-1$ expression to the $\beta \to 0$ limit of the slow-roll case only if one assumes that the constant-roll condition still holds at $\beta=0$ and one has $\epsilon_2 = 2\epsilon_1 -6 \beta$. When $\beta=0$, we have $\epsilon_2=2\epsilon_1$, and using this result in Eq.~(\ref{nsm1s}) one gets $n_s-1 \simeq -4\epsilon_1$ in the $\beta \to 0$ limit. In a sense, the spectral index of slow-roll inflation matches with that of CRI in the zero $\beta$ limit if we assume the system to be constrained by the constant-roll condition in that limit. Thus \(n_s-1\simeq-4\epsilon_1\) is not generic slow-roll; it is the slow-roll spectral formula evaluated on the special \(\beta=0\) constant-roll trajectory.  

For the main reconstructed branches considered here, the strict
$\beta\to0$ limit does not give a generic rolling slow-roll potential. Rather,
the potential degenerates to an exactly flat de Sitter form,
\begin{equation}
    V(\phi)\longrightarrow 3M^2M_{\rm Pl}^2,
    \qquad
    H(\phi)\longrightarrow M,
    \qquad
    \dot\phi\longrightarrow 0 .
\end{equation}
Therefore, in the background sector, the $\beta\to0$ limit is an exact
de Sitter limit of the constant-roll branch, not a generic slow-roll
potential limit. This shows that there are two distinct correspondences: Apparently
\begin{equation}
    \beta\to0
    \quad\Longrightarrow\quad
    n_s\to n_s^{\rm slow\text{-}roll}
\end{equation}
at the level of the scalar spectrum, although this limit has its own subtlety, we also have:
\begin{equation}
    \beta\to0
    \quad\Longrightarrow\quad
    V(\phi)\to \text{constant}
\end{equation}
at the level of the reconstructed background potential.

Thus, although the constant-roll construction is spectrally connected to slow-roll
inflation via a nontrivial limiting process, but dynamically it approaches an exact de Sitter configuration.
The same slow-roll value of the scalar spectral index arise
from a limiting constant-roll branch whose background dynamics is not the
usual rolling slow-roll dynamics. This distinction reflects the fact that
constant-roll and slow-roll impose different restrictions on the inflationary
background.

Strictly speaking, the phrase “slow-roll limit of constant-roll inflation” is potentially misleading. The limit \(\beta\to0\) produces a slow-roll-like expression for the scalar spectral index, but the reconstructed constant-roll background does not become a generic slow-roll inflationary model. For the branches considered here, the potential degenerates to an exactly flat de Sitter form and the scalar dynamics freezes. Thus the correspondence with slow roll is apparently at the spectral or phenomenological level, but it is not a regular perturbative correspondence at the level of the background dynamics.

This makes the result conceptually more interesting. Constant-roll inflation should not be viewed merely as a perturbative deformation of conventional slow-roll inflation. Although the scalar spectral index has a slow-roll-like limiting form as \(\beta\to0\), the reconstructed background dynamics does not reduce to a generic slow-roll potential. Thus CRI represents a dynamically distinct inflationary realization whose observational predictions can nevertheless mimic, and in the present analysis marginally improve upon, the standard slow-roll description.
\section{Conclusion}
\label{conclu}

In this work we have studied constant-roll inflation as a controlled extension of the standard slow-roll framework and confronted it with current CMB observations. The central idea is to treat the constant-roll parameter $\beta$ as a primary inflationary parameter, rather than introducing the scalar spectral index merely as a phenomenological input. In this formulation, $\beta$ controls the departure from the slow-roll attractor, determines the structure of the inflationary background, classifies the possible branches of the reconstructed potential, and fixes the scalar tilt once the number of $e$-folds is specified.

Using the Hamilton--Jacobi formulation, we showed that imposing the constant-roll condition strongly restricts the allowed form of the Hubble parameter as a function of the inflaton field. The sign of $\beta$ naturally separates the theory into different branches. For positive $\beta$, the solutions are hyperbolic, while for negative $\beta$, after writing $\beta=-\widetilde\beta$, the solutions become oscillatory. The scalar potentials are then reconstructed from the Hamilton--Jacobi relation. Thus constant-roll inflation is not an arbitrary deformation of slow roll; it defines a constrained class of inflationary backgrounds.

We then derived the scalar perturbation spectrum for the constant-roll background. The Mukhanov--Sasaki equation itself has the standard form for canonical single-field inflation, while the constant-roll condition enters through the background-dependent quantity $z''/z$. This leads to a relation between the scalar spectral index, the constant-roll parameter, and the number of $e$-folds before the end of inflation. For a representative value of sixty $e$-folds, the observed scalar tilt admits two small $|\beta|$ solutions, one with negative $\beta$ and one with positive $\beta$. This two-branch structure is one of the main theoretical features of the model.

The observational analysis shows that current CMB data constrain $\beta$ to be small, but do not require it to vanish. The Planck 2018 analysis gives a bimodal posterior distribution for $\beta$, reflecting the two analytical branches of the constant-roll solution. The two branches correspond to different reconstructed potentials, but they produce very similar scalar spectral indices over the observed CMB window. Consequently, present data cannot decisively distinguish between them. The Planck-only analysis gives a 95\% bound of approximately $|\beta|<0.006)$, while the combined Planck 2018 plus ACT DR6 analysis tightens this to approximately $(|\beta|<0.0046)$. An important aspect of the analysis is that $\beta$ replaces $n_s$ as the primordial-shape parameter. Therefore, for fixed (N), the constant-roll model has the same number of sampled cosmological parameters as the standard six-parameter fit. The model is not obtaining agreement with the data by adding an extra degree of freedom. Rather, it trades the phenomenological scalar tilt for a dynamical inflationary parameter. This makes the comparison with the standard slow-roll-inspired power-law spectrum especially meaningful.

The best-fit comparison suggests that the constant-roll parametrization gives a fit that is statistically comparable to, and marginally better than, the standard reference model. However, the improvement in the minimum $\chi^2$ is small and should not be interpreted as decisive evidence in favor of constant roll. The correct conclusion is more conservative but still significant: constant-roll inflation is observationally competitive with the standard slow-roll description, and current CMB data allow small constant-roll departures from the exact slow-roll limit. The slow-roll point remains fully consistent with the data, but it is not uniquely selected by them. The manuscript reports a small improvement, $\Delta\chi^2\simeq -0.86$, which supports only a marginal improvement at equal parameter count.

We also estimated the characteristic energy scale associated with the allowed constant-roll branches using the observed scalar amplitude. Both the positive- and negative $\beta$ branches lead to an inflationary scale of order $10^{-5}M_{\rm Pl}$. This indicates that the two branches are not only degenerate at the level of the scalar tilt, but also lead to comparable inflationary normalizations. This supports the interpretation that the branch structure is a genuine feature of the constant-roll dynamics rather than an artifact of the amplitude normalization.

The main conclusion of this work is therefore that constant-roll inflation provides a viable and competitive alternative to the standard slow-roll parametrization of primordial perturbations. The constant-roll parameter $\beta$ can be regarded as a fundamental inflationary parameter that is directly constrained by CMB data. Current observations force $\beta$ to be small, but they do not force it to be exactly zero. The resulting two-branch degeneracy shows that distinct constant-roll potentials can remain observationally indistinguishable at the level of present scalar CMB constraints.

\begin{acknowledgments}
We acknowledge the use of the \texttt{Cobaya} code for Bayesian parameter estimation~\cite{Torrado:2021qmv}.
\end{acknowledgments}

\bibliographystyle{unsrtnat}
\bibliography{references}

@article{Kazanas:1980tx,
    author = "Kazanas, D.",
    title = "{Dynamics of the Universe and Spontaneous Symmetry Breaking}",
    doi = "10.1086/183361",
    journal = "Astrophys. J. Lett.",
    volume = "241",
    pages = "L59--L63",
    year = "1980"
}

@article{Guth:1980zm,
    author = "Guth, Alan H.",
    editor = "Fang, Li-Zhi and Ruffini, R.",
    title = "{The Inflationary Universe: A Possible Solution to the Horizon and Flatness Problems}",
    reportNumber = "SLAC-PUB-2576",
    doi = "10.1103/PhysRevD.23.347",
    journal = "Phys. Rev. D",
    volume = "23",
    pages = "347--356",
    year = "1981"
}

@article{Sato:1981ds,
    author = "Sato, K.",
    editor = "Fang, Li-Zhi and Ruffini, R.",
    title = "{Cosmological Baryon Number Domain Structure and the First Order Phase Transition of a Vacuum}",
    doi = "10.1016/0370-2693(81)90805-4",
    journal = "Phys. Lett. B",
    volume = "99",
    pages = "66--70",
    year = "1981"
}

@article{Sato:1980yn,
    author = "Sato, K.",
    title = "{First Order Phase Transition of a Vacuum and Expansion of the Universe}",
    reportNumber = "NORDITA-80-29",
    journal = "Mon. Not. Roy. Astron. Soc.",
    volume = "195",
    pages = "467--479",
    year = "1981"
}

@article{Linde:1981mu,
    author = "Linde, Andrei D.",
    editor = "Fang, Li-Zhi and Ruffini, R.",
    title = "{A New Inflationary Universe Scenario: A Possible Solution of the Horizon, Flatness, Homogeneity, Isotropy and Primordial Monopole Problems}",
    reportNumber = "LEBEDEV-81-229",
    doi = "10.1016/0370-2693(82)91219-9",
    journal = "Phys. Lett. B",
    volume = "108",
    pages = "389--393",
    year = "1982"
}

@article{Carrasco_2015,
   title={Cosmological attractors and initial conditions for inflation},
   volume={92},
   ISSN={1550-2368},
   url={http://dx.doi.org/10.1103/PhysRevD.92.063519},
   DOI={10.1103/physrevd.92.063519},
   number={6},
   journal={Physical Review D},
   publisher={American Physical Society (APS)},
   author={Carrasco, John Joseph M. and Kallosh, Renata and Linde, Andrei},
   year={2015},
   month=sep }

@article{Riotto:2002yw,
    author = "Riotto, Antonio",
    editor = "Dvali, G. and Perez-Lorenzana, Abdel and Senjanovic, G. and Thompson, G. and Vissani, F.",
    title = "{Inflation and the theory of cosmological perturbations}",
    eprint = "hep-ph/0210162",
    archivePrefix = "arXiv",
    reportNumber = "DFPD-TH-02-22",
    journal = "ICTP Lect. Notes Ser.",
    volume = "14",
    pages = "317--413",
    year = "2003"
}

@article{Martin:2012pe,
    author = "Martin, Jerome and Motohashi, Hayato and Suyama, Teruaki",
    title = "{Ultra Slow-Roll Inflation and the non-Gaussianity Consistency Relation}",
    eprint = "1211.0083",
    archivePrefix = "arXiv",
    primaryClass = "astro-ph.CO",
    reportNumber = "RESCEU-47-12",
    doi = "10.1103/PhysRevD.87.023514",
    journal = "Phys. Rev. D",
    volume = "87",
    number = "2",
    pages = "023514",
    year = "2013"
}

@misc{brandenberger1993classicalquantumtheoryperturbations,
      title={Classical and Quantum Theory of Perturbations in Inflationary Universe Models}, 
      author={R. Brandenberger and H. Feldman and V. Mukhanov},
      year={1993},
      eprint={astro-ph/9307016},
      archivePrefix={arXiv},
      primaryClass={astro-ph},
      url={https://arxiv.org/abs/astro-ph/9307016}, 
}

@article{Mukhanov:1981xt,
    author = "Mukhanov, Viatcheslav F. and Chibisov, G. V.",
    title = "{Quantum Fluctuations and a Nonsingular Universe}",
    journal = "JETP Lett.",
    volume = "33",
    pages = "532--535",
    year = "1981"
}

@article{Guth:1982ec,
    author = "Guth, Alan H. and Pi, S. Y.",
    title = "{Fluctuations in the New Inflationary Universe}",
    doi = "10.1103/PhysRevLett.49.1110",
    journal = "Phys. Rev. Lett.",
    volume = "49",
    pages = "1110--1113",
    year = "1982"
}

@article{Hawking:1982cz,
    author = "Hawking, S. W.",
    title = "{The Development of Irregularities in a Single Bubble Inflationary Universe}",
    reportNumber = "Print-83-0015 (CAMBRIDGE)",
    doi = "10.1016/0370-2693(82)90373-2",
    journal = "Phys. Lett. B",
    volume = "115",
    pages = "295",
    year = "1982"
}

@article{Bardeen:1983qw,
    author = "Bardeen, James M. and Steinhardt, Paul J. and Turner, Michael S.",
    title = "{Spontaneous Creation of Almost Scale - Free Density Perturbations in an Inflationary Universe}",
    reportNumber = "UPR-0202T, EFI-83-13-CHICAGO",
    doi = "10.1103/PhysRevD.28.679",
    journal = "Phys. Rev. D",
    volume = "28",
    pages = "679",
    year = "1983"
}

@article{Mukhanov:1985rz,
    author = "Mukhanov, Viatcheslav F.",
    title = "{Gravitational Instability of the Universe Filled with a Scalar Field}",
    journal = "JETP Lett.",
    volume = "41",
    pages = "493--496",
    year = "1985"
}

@article{Sasaki:1986hm,
    author = "Sasaki, Misao",
    title = "{Large Scale Quantum Fluctuations in the Inflationary Universe}",
    reportNumber = "RRK-86-29",
    doi = "10.1143/PTP.76.1036",
    journal = "Prog. Theor. Phys.",
    volume = "76",
    pages = "1036",
    year = "1986"
}

@article{Motohashi_2015,
   title={Inflation with a constant rate of roll},
   volume={2015},
   ISSN={1475-7516},
   url={http://dx.doi.org/10.1088/1475-7516/2015/09/018},
   DOI={10.1088/1475-7516/2015/09/018},
   number={09},
   journal={Journal of Cosmology and Astroparticle Physics},
   publisher={IOP Publishing},
   author={Motohashi, Hayato and Starobinsky, Alexei A. and Yokoyama, Jun’ichi},
   year={2015},
   month=sep, pages={018–018} }

@article{Yi:2017mxs,
    author = "Yi, Zhu and Gong, Yungui",
    title = "{On the constant-roll inflation}",
    eprint = "1712.07478",
    archivePrefix = "arXiv",
    primaryClass = "gr-qc",
    doi = "10.1088/1475-7516/2018/03/052",
    journal = "JCAP",
    volume = "03",
    pages = "052",
    year = "2018"
}

@article{Guerrero:2020lng,
    author = "Guerrero, Merce and Rubiera-Garcia, Diego and Saez-Chillon Gomez, Diego",
    title = "{Constant roll inflation in multifield models}",
    eprint = "2008.07260",
    archivePrefix = "arXiv",
    primaryClass = "gr-qc",
    doi = "10.1103/PhysRevD.102.123528",
    journal = "Phys. Rev. D",
    volume = "102",
    pages = "123528",
    year = "2020"
}

@article{Motohashi_2017,
   title={f(R) constant-roll inflation},
   volume={77},
   ISSN={1434-6052},
   url={http://dx.doi.org/10.1140/epjc/s10052-017-5109-x},
   DOI={10.1140/epjc/s10052-017-5109-x},
   number={8},
   journal={The European Physical Journal C},
   publisher={Springer Science and Business Media LLC},
   author={Motohashi, Hayato and Starobinsky, Alexei A.},
   year={2017},
   month=aug }

@article{Motohashi_20171,
   title={Constant-roll inflation: Confrontation with recent observational data},
   volume={117},
   ISSN={1286-4854},
   url={http://dx.doi.org/10.1209/0295-5075/117/39001},
   DOI={10.1209/0295-5075/117/39001},
   number={3},
   journal={EPL (Europhysics Letters)},
   publisher={IOP Publishing},
   author={Motohashi, Hayato and Starobinsky, Alexei A.},
   year={2017},
   month=feb, pages={39001} }

@article{Motohashi_2017pbh,
   title={Primordial black holes and slow-roll violation},
   volume={96},
   ISSN={2470-0029},
   url={http://dx.doi.org/10.1103/PhysRevD.96.063503},
   DOI={10.1103/physrevd.96.063503},
   number={6},
   journal={Physical Review D},
   publisher={American Physical Society (APS)},
   author={Motohashi, Hayato and Hu, Wayne},
   year={2017},
   month=sep }

@article{Gao_2017,
   title={Reconstruction of constant slow-roll inflation},
   volume={60},
   ISSN={1869-1927},
   url={http://dx.doi.org/10.1007/s11433-017-9065-4},
   DOI={10.1007/s11433-017-9065-4},
   number={9},
   journal={Science China Physics, Mechanics \&amp; Astronomy},
   publisher={Springer Science and Business Media LLC},
   author={Gao, Qing},
   year={2017},
   month=jul }

@article{Odintsov_2017,
   title={Inflation with a smooth constant-roll to constant-roll era transition},
   volume={96},
   ISSN={2470-0029},
   url={http://dx.doi.org/10.1103/PhysRevD.96.024029},
   DOI={10.1103/physrevd.96.024029},
   number={2},
   journal={Physical Review D},
   publisher={American Physical Society (APS)},
   author={Odintsov, S.D. and Oikonomou, V.K.},
   year={2017},
   month=jul }

@article{Ito_2018,
   title={Anisotropic constant-roll Inflation},
   volume={78},
   ISSN={1434-6052},
   url={http://dx.doi.org/10.1140/epjc/s10052-018-5534-5},
   DOI={10.1140/epjc/s10052-018-5534-5},
   number={1},
   journal={The European Physical Journal C},
   publisher={Springer Science and Business Media LLC},
   author={Ito, Asuka and Soda, Jiro},
   year={2018},
   month=jan }

@article{Karam_2018,
   title={Constant-roll (quasi-)linear inflation},
   volume={2018},
   ISSN={1475-7516},
   url={http://dx.doi.org/10.1088/1475-7516/2018/05/011},
   DOI={10.1088/1475-7516/2018/05/011},
   number={05},
   journal={Journal of Cosmology and Astroparticle Physics},
   publisher={IOP Publishing},
   author={Karam, A. and Marzola, L. and Pappas, T. and Racioppi, A. and Tamvakis, K.},
   year={2018},
   month=may, pages={011–011} }

@article{Cicciarella_2018,
   title={New perspectives on constant-roll inflation},
   volume={2018},
   ISSN={1475-7516},
   url={http://dx.doi.org/10.1088/1475-7516/2018/01/024},
   DOI={10.1088/1475-7516/2018/01/024},
   number={01},
   journal={Journal of Cosmology and Astroparticle Physics},
   publisher={IOP Publishing},
   author={Cicciarella, Francesco and Mabillard, Joel and Pieroni, Mauro},
   year={2018},
   month=jan, pages={024–024} }

@article{Anguelova_2018,
   title={Systematics of constant roll inflation},
   volume={2018},
   ISSN={1475-7516},
   url={http://dx.doi.org/10.1088/1475-7516/2018/02/004},
   DOI={10.1088/1475-7516/2018/02/004},
   number={02},
   journal={Journal of Cosmology and Astroparticle Physics},
   publisher={IOP Publishing},
   author={Anguelova, Lilia and Suranyi, Peter and Wijewardhana, L.C.R.},
   year={2018},
   month=feb, pages={004–004} }

@article{Gao_2018,
   title={The observational constraint on constant-roll inflation},
   volume={61},
   ISSN={1869-1927},
   url={http://dx.doi.org/10.1007/s11433-018-9197-2},
   DOI={10.1007/s11433-018-9197-2},
   number={7},
   journal={Science China Physics, Mechanics \&amp; Astronomy},
   publisher={Springer Science and Business Media LLC},
   author={Gao, Qing},
   year={2018},
   month=mar }

@article{Gao_2019,
   title={On the Constant-Roll Inflation with Large and Small \(\eta H\)},
   volume={5},
   ISSN={2218-1997},
   url={http://dx.doi.org/10.3390/universe5110215},
   DOI={10.3390/universe5110215},
   number={11},
   journal={Universe},
   publisher={MDPI AG},
   author={Gao, Qing and Gong, Yungui and Yi, Zhu},
   year={2019},
   month=oct, pages={215} }

@article{Odintsov_2017_02,
   title={Unification of constant-roll inflation and dark energy with logarithmic R2-corrected and exponential F(R) gravity},
   volume={923},
   ISSN={0550-3213},
   url={http://dx.doi.org/10.1016/j.nuclphysb.2017.08.018},
   DOI={10.1016/j.nuclphysb.2017.08.018},
   journal={Nuclear Physics B},
   publisher={Elsevier BV},
   author={Odintsov, S.D. and Oikonomou, V.K. and Sebastiani, L.},
   year={2017},
   month=oct, pages={608–632} }

@article{Mun_2021,
   title={Constant-roll warm inflation and the \(\beta\) function approach},
   volume={103},
   ISSN={2470-0029},
   url={http://dx.doi.org/10.1103/PhysRevD.103.083527},
   DOI={10.1103/physrevd.103.083527},
   number={8},
   journal={Physical Review D},
   publisher={American Physical Society (APS)},
   author={Mun, Ui Ri},
   year={2021},
   month=apr }

@article{Biswas:2025vlz,
    author  = {Biswas, Sandip and Hussain, Saddam and Bhattacharya, Kaushik},
  title   = {Dynamical Systems Approach to Non-Slow-Roll Inflationary Models},
  journal = {General Relativity and Gravitation},
  year    = {2026},
  volume  = {58},
  number  = {2},
  pages   = {14},
  doi     = {10.1007/s10714-026-03513-6},
  url     = {https://doi.org/10.1007/s10714-026-03513-6},
}

@article{Torrado:2021qmv,
    author = "Torrado, Jesus and Lewis, Antony",
    title = "{Cobaya: Code for Bayesian Analysis of hierarchical physical models}",
    eprint = "2005.05290",
    archivePrefix = "arXiv",
    primaryClass = "astro-ph.IM",
    doi = "10.1088/1475-7516/2021/05/057",
    journal = "JCAP",
    volume = "05",
    pages = "057",
    year = "2021"
}

@article{Aghanim2020,
  author  = {Aghanim, N. and Akrami, Y. and Ashdown, M. and others},
  title   = {Planck 2018 results. VI. Cosmological parameters},
  journal = {Astronomy \& Astrophysics},
  volume  = {641},
  pages   = {A6},
  year    = {2020},
  doi     = {10.1051/0004-6361/201833910}
}

@article{Leach_2002,
   title={Cosmological parameter estimation and the inflationary cosmology},
   volume={66},
   ISSN={1089-4918},
   url={http://dx.doi.org/10.1103/PhysRevD.66.023515},
   DOI={10.1103/physrevd.66.023515},
   number={2},
   journal={Physical Review D},
   publisher={American Physical Society (APS)},
   author={Leach, Samuel M. and Liddle, Andrew R. and Martin, Jérôme and Schwarz, Dominik J.},
   year={2002},
   month=July }

@article{Mortonson_2011,
   title={Bayesian analysis of inflation: Parameter estimation for single field models},
   volume={83},
   ISSN={1550-2368},
   url={http://dx.doi.org/10.1103/PhysRevD.83.043505},
   DOI={10.1103/physrevd.83.043505},
   number={4},
   journal={Physical Review D},
   publisher={American Physical Society (APS)},
   author={Mortonson, Michael J. and Peiris, Hiranya V. and Easther, Richard},
   year={2011},
   month=Feb }

@article{Easther_2012,
   title={Bayesian analysis of inflation. II. Model selection and constraints on reheating},
   volume={85},
   ISSN={1550-2368},
   url={http://dx.doi.org/10.1103/PhysRevD.85.103533},
   DOI={10.1103/physrevd.85.103533},
   number={10},
   journal={Physical Review D},
   publisher={American Physical Society (APS)},
   author={Easther, Richard and Peiris, Hiranya V.},
   year={2012},
   month=May }

@misc{abazajian2019cmbs4sciencecasereference,
  title={CMB-S4 Science Case, Reference Design, and Project Plan},
  author={Abazajian et al.},
  year={2019},
  eprint={1907.04473},
  archivePrefix={arXiv},
  primaryClass={astro-ph.IM},
  url={https://arxiv.org/abs/1907.04473}
}

@article{Matsumura_2014,
   title={Mission Design of LiteBIRD},
   volume={176},
   ISSN={1573-7357},
   url={http://dx.doi.org/10.1007/s10909-013-0996-1},
   DOI={10.1007/s10909-013-0996-1},
   number={5-6},
   journal={Journal of Low Temperature Physics},
   publisher={Springer Science and Business Media LLC},
   author={Matsumura et al.
},
   year={2014},
   month=Jan, pages={733–740} }

@ARTICLE{CobeCosmo1992,
       author = {{Wright}, E.~L. and {Meyer}, S.~S. and {Bennett}, C.~L. and {Boggess}, A. and {Cheng}, E.~S. and {Hauser}, M.~G. and {Kogut}, A. and {Lineweaver}, C. and {Mather}, J.~C. and {Smoot}, G.~F. and {Weiss}, R. and {Gulkis}, S. and {Hinshaw}, G. and {Janssen}, M. and , K. and {Lubin}, P.~M. and {Moseley}, S.~H., Jr. and {Murdock}, T.~L. and {Shafer}, R.~A. and {Silverberg}, R.~F. and {Wilkinson}, D.~T.},
        title = "{Interpretation of the Cosmic Microwave Background Radiation Anisotropy Detected by the COBE Differential Microwave Radiometer}",
      journal = {\textsl{The Astrophysical Journal Letters}},
         year = 1992,
        month = sep,
       volume = {396},
        pages = {L13},
          doi = {10.1086/186506},
       adsurl = {https://harvard.edu}
}

@ARTICLE{WMAPCosmo2013,
       author = {{Hinshaw}, G. and {Larson}, D. and {Komatsu}, E. and {Spergel}, D.~N. and {Bennett}, C.~L. and {Dunkley}, J. and {Nolta}, M.~R. and {Halpern}, M. and {Hill}, R.~S. and {Jarosik}, N. and {Kogut}, A. and {Limon}, M. and {Meyer}, S.~S. and {Odegard}, N. and {Page}, L. and {Smith}, K.~M. and {Weiland}, J.~L. and {Wollack}, E. and {Wright}, E.~L.},
        title = "{Nine-Year Wilkinson Microwave Anisotropy Probe (WMAP) Observations: Cosmological Parameter Results}",
      journal = {\textsl{The Astrophysical Journal Supplement Series}},
         year = 2013,
        month = oct,
       volume = {208},
        number = {2},
        eid = {19},
        pages = {19},
          doi = {10.1088/0067-0049/208/2/19},
archivePrefix = {arXiv},
       eprint = {1212.5226},
 primaryClass = {astro-ph.CO},
       adsurl = {https://harvard.edu}
}

@ARTICLE{PlanckCosmo2020,
       author = {{Planck Collaboration} and {Aghanim}, N. and {Akrami}, Y. and {Ashdown}, M. and {Aumont}, J. and {Baccigalupi}, C. and {Ballardini}, M. and {Banday}, A.~J. and {Barreiro}, R.~B. and {Bartolo}, N. and {Basak}, S. and {Battye}, R. and {Benabed}, K. and {Bernard}, J. -P. and {Bersanelli}, M. and {Bielewicz}, P. and {Bock}, J.~J. and {Bond}, J.~R. and {Borrill}, J. and {Bouchet}, F.~R. and {Boulanger}, F. and {Bucher}, M. and {Burigana}, C. and {Butler}, R.~C. and {Calabrese}, E. and {Cardoso}, J. -F. and {Carron}, J. and {Challinor}, A. and {Chiang}, H.~C. and {Chluba}, J. and {Colombo}, L.~P.~L. and {Combet}, C. and {Contreras}, D. and {Crill}, B.~P. and {Cuttaia}, F. and {de Bernardis}, P. and {de Zotti}, G. and {Delabrouille}, J. and {Delouis}, J. -M. and {Di Valentino}, E. and {Diego}, J.~M. and {Dor{\'e}}, O. and {Douspis}, M. and {Ducout}, A. and {Dupac}, X. and {Dusini}, S. and {Efstathiou}, G. and {Elsner}, F. and {En{\ss}lin}, T.~A. and {Eriksen}, H.~K. and {Fantaye}, Y. and {Farhang}, M. and {Fergusson}, J. and {Fernandez-Cobos}, R. and {Finelli}, F. and {Forastieri}, F. and {Frailis}, M. and {Fraisse}, A.~A. and {Franceschi}, E. and {Frolov}, A. and {Galeotta}, S. and {Galli}, S. and {Ganga}, K. and {G{\'e}nova-Santos}, R.~T. and {Gerbino}, M. and {Ghosh}, T. and {Gonzalez-Nuevo}, J. and {G{\'o}rski}, K.~M. and {Gratton}, S. and {Gruppuso}, A. and {Gudmundsson}, J.~E. and {Hamann}, J. and {Handley}, W. and {Hansen}, F.~K. and {Herranz}, D. and {Hildebrandt}, S. and {Hivon}, E. and {Huang}, Z. and {Jaffe}, A.~H. and {Jones}, W.~C. and {Karakci}, A. and {Keih{\"a}nen}, E. and {Keskitalo}, R. and {Kiiveri}, K. and {Kim}, J. and {Kisner}, T.~S. and {Knox}, L. and {Krachmalnicoff}, N. and {Kunz}, M. and {Kurki-Suonio}, H. and {Lagache}, G. and {Lamarre}, J. -M. and {Lasenby}, A. and {Lattanzi}, M. and {Lawrence}, C.~R. and {Le Jeune}, M. and {Lemos}, P. and {Lesgourgues}, J. and {Levrier}, F. and {Lewis}, A. and {Liguori}, M. and {Lilje}, P.~B. and {Lilley}, M. and {Lindholm}, V. and {L{\'o}pez-Caniego}, M. and {Lubin}, P.~M. and {Ma}, Y. -Z. and {Mac{\'\i}as-P{\'e}rez}, J.~F. and {Maggio}, G. and {Maino}, D. and {Mandolesi}, N. and {Mangilli}, A. and {Marcos-Caballero}, A. and {Marinucci}, D. and {Mart{\'\i}nez-Gonz{\'a}lez}, E. and {Masi}, S. and {Matarrese}, S. and {Mauri}, N. and {McEwen}, J.~D. and {Meinhold}, P.~R. and {Melchiorri}, A. and {Mennella}, A. and {Migliaccio}, M. and {Millea}, M. and {Mitra}, S. and {Miville-Desch{\^e}nes}, M. -A. and {Molinari}, D. and {Montier}, L. and {Morgante}, G. and {Moss}, A. and {Natoli}, P. and {N{\o}rgaard-Nielsen}, H.~U. and {Pagano}, L. and {Paoletti}, D. and {Partridge}, B. and {Patanchon}, G. and {Peiris}, H.~V. and {Perrotta}, F. and {Pettorino}, V. and {Piacentini}, F. and {Puget}, J. -L. and {Rachen}, J.~P. and {Reinecke}, M. and {Remazeilles}, M. and {Renzi}, A. and {Rocha}, G. and {Rosset}, C. and {Roudier}, G. and {Rubi{\~n}o-Mart{\'\i}n}, J.~A. and {Ruiz-Granados}, B. and {Salvati}, L. and {Sandri}, M. and {Savelainen}, M. and {Scott}, D. and {Shellard}, E.~P.~S. and {Sirignano}, C. and {Sirri}, G. and {Spencer}, L.~D. and {Sunyaev}, R. and {Suur-Uski}, A. -S. and {Tauber}, J.~A. and {Tavagnacco}, D. and {Tenti}, M. and {Toffolatti}, L. and {Tomasi}, M. and {Trombetti}, T. and {Valenziano}, L. and {Valiviita}, J. and {Van Tent}, B. and {Vibert}, L. and {Vielva}, P. and {Villa}, F. and {Vittorio}, N. and {Wandelt}, B.~D. and {Wehus}, I.~K. and {White}, M. and {White}, S.~D.~M. and {Zacchei}, A. and {Zonca}, A.},
        title = "{Planck 2018 results. VI. Cosmological parameters}",
      journal = {\textsl{Astronomy \& Astrophysics}},
         year = 2020,
        month = sep,
       volume = {641},
          eid = {A6},
        pages = {A6},
          doi = {10.1051/0004-6361/201833910},
archivePrefix = {arXiv},
       eprint = {1807.06209},
 primaryClass = {astro-ph.CO},
       adsurl = {https://harvard.edu}
}

@ARTICLE{ACTCosmo2025,
       author = {{Louis}, Thibaut and {Abitbol}, Maximilian H. and {Adame}, Maria and {Ade}, Peter A.~R. and {Aguilar}, Fausto and {Aiola}, Simone and {Alvarez}, Marisa and {Amiri}, Mandana and {Arnold}, Kam and {Ashton}, Peter and et al.},
        title = "{The Atacama Cosmology Telescope: DR6 Power Spectra, Likelihoods and $\Lambda$CDM Parameters}",
      journal = {\textsl{Journal of Cosmology and Astroparticle Physics}},
         year = 2025,
        month = nov,
       volume = {2025},
       number = {11},
          eid = {062},
        pages = {062},
          doi = {10.1088/1475-7516/2025/11/062},
archivePrefix = {arXiv},
       eprint = {2503.14452},
 primaryClass = {astro-ph.CO},
       adsurl = {https://harvard.edu}
}

@article{MNRAS_Saha1,
    author = {Joseph, Albin and Purkayastha, Ujjal and Saha, Rajib},
    title = {A foreground model-independent Bayesian CMB temperature and polarization signal reconstruction and cosmological parameter estimation over large angular scales},
    journal = {Monthly Notices of the Royal Astronomical Society},
    volume = {520},
    number = {1},
    pages = {976-987},
    year = {2023},
    month = {03},
    issn = {0035-8711},
    doi = {10.1093/mnras/stad187},
    url = {https://doi.org/10.1093/mnras/stad187},
    eprint = {https://academic.oup.com/mnras/article-pdf/520/1/976/49058071/stad187.pdf},
}

@ARTICLE{MNRAS_Saha2,
       author = {{Joseph}, Albin and {Saha}, Rajib},
        title = "{Forecast analysis on interacting dark energy models from future generation PICO and DESI missions}",
      journal = {MNRAS},
         year = 2023,
        month = feb,
       volume = {519},
       number = {2},
        pages = {1809-1822},
          doi = {10.1093/mnras/stac3586},
archivePrefix = {arXiv},
       eprint = {2209.07167},
 primaryClass = {astro-ph.CO},
       adsurl = {https://ui.adsabs.harvard.edu/abs/2023MNRAS.519.1809J}
}

@ARTICLE{ApJ_Saha,
       author = {{Yadav}, Sarvesh Kumar and {Saha}, Rajib},
        title = "{A Bayesian ILC Method for CMB B-mode Posterior Estimation and Reconstruction of Primordial Gravity Wave Signal}",
      journal = {ApJ},
         year = 2021,
        month = jun,
       volume = {914},
       number = {2},
          eid = {119},
        pages = {119},
          doi = {10.3847/1538-4357/abfd9b},
archivePrefix = {arXiv},
       eprint = {2009.14567},
 primaryClass = {astro-ph.CO},
       adsurl = {https://ui.adsabs.harvard.edu/abs/2021ApJ...914..119Y}
}

@ARTICLE{Saha2022,
       author = {{Joseph}, Albin and {Saha}, Rajib},
        title = "{Dark energy with oscillatory tracking potential: observational constraints and perturbative effects}",
      journal = {MNRAS},
         year = 2022,
        month = apr,
       volume = {511},
       number = {2},
        pages = {1637-1646},
          doi = {10.1093/mnras/stac201},
archivePrefix = {arXiv},
       eprint = {2110.00229},
 primaryClass = {astro-ph.CO},
       adsurl = {https://ui.adsabs.harvard.edu/abs/2022MNRAS.511.1637J}
}

@ARTICLE{SahaSrikanta2022,
       author = {{Pal}, Srikanta and {Saha}, Rajib},
        title = "{On direct estimation of density parameters and Hubble constant for {\ensuremath{\Lambda}}CDM universe using Hubble measurements}",
      journal = {Physica Scripta},
         year = 2024,
        month = aug,
       volume = {99},
       number = {8},
          eid = {085025},
        pages = {085025},
          doi = {10.1088/1402-4896/ad5f5b},
archivePrefix = {arXiv},
       eprint = {2204.07099},
 primaryClass = {astro-ph.CO},
       adsurl = {https://ui.adsabs.harvard.edu/abs/2024PhyS...99h5025P}
}

@ARTICLE{SahaSrikanta2024,
       author = {{Pal}, Srikanta and {Saha}, Rajib},
        title = "{ParamANN: a neural network to estimate cosmological parameters for {\ensuremath{\Lambda}}CDM Universe using Hubble measurements}",
      journal = {Physica Scripta},
         year = 2024,
        month = nov,
       volume = {99},
       number = {11},
          eid = {115007},
        pages = {115007},
          doi = {10.1088/1402-4896/ad804d},
archivePrefix = {arXiv},
       eprint = {2309.15179},
 primaryClass = {astro-ph.CO},
       adsurl = {https://ui.adsabs.harvard.edu/abs/2024PhyS...99k5007P}
}

\end{document}